\documentstyle[12pt]{article}

\newcommand {\nn}    {\nonumber}
\newcommand {\vs}[1]  { \vspace*{#1 cm} }

\newcounter{eq}
\newcounter{sc}

\newcommand {\MPL}  {Mod.Phys.Lett.}
\newcommand {\NP}   {Nucl.Phys.}
\newcommand {\PL}   {Phys.Lett.}
\newcommand {\PR}   {Phys.Rev.}
\newcommand {\PRL}   {Phys.Rev.Lett.}

\newcommand {\AP}   {Ann.of Phys.}

\newcommand {\JHEP}  {J.High Energy Phys.}

\def\overleftrightarrow#1{\vbox{\ialign{##\crcr
 $\leftrightarrow$\crcr\noalign{\kern-1pt\nointerlineskip}
 $\hfil\displaystyle{#1}\hfil$\crcr}}}

\newlength{\minitwocolumn}
\begin{document}

\begin{flushright}
CHIBA-EP-113\\
EDO-EP-26\\
April, 1999\\
\end{flushright}
\vspace{20pt}

\pagestyle{empty}
\baselineskip15pt

\begin{center}
{\large\bf Duality of Super D-brane Actions \\
in \\
General Type II Supergravity Background
 \vskip 1mm
}

\vspace{10mm}

Tadahiko Kimura
          \footnote{
          E-mail address:\ kimura@cuphd.nd.chiba-u.ac.jp
                  }
\\

\vspace{5mm}
          Department of Physics, Faculty of Science,
          Chiba University, Chiba 263-8522, JAPAN \\

\vspace{5mm}

and

\vspace{5mm}

Ichiro Oda
          \footnote{
          E-mail address:\ ioda@edogawa-u.ac.jp
                  }
\\
\vspace{5mm}
          Edogawa University,
          474 Komaki, Nagareyama City, Chiba 270-0198, JAPAN \\

\end{center}


\vspace{5mm}
\begin{abstract}
We examine duality transformations of supersymmetric and
$\kappa$-symmetric Dp-brane actions in a general type II
supergravity background where in particular the dilaton
and the axion are supposed to not be zero or a constant
but a general superfield.
Due to non-constant dilaton and axion, we can explicitly
show that the dilaton and the axion as well as the two
2-form gauge potentials transform as doublets under the
$SL(2,R)$ transformation from the point of view of the
world-volume field theory.
\vspace{3mm}
\begin{flushleft}
{\it PACS:} 11-25.-w; 12.60.Jv; 04.65 \\
{\it Key words:} D-branes; Duality; Supergravity; Dilaton; Axion
\end{flushleft}

\end{abstract}

\newpage
\pagestyle{plain}
\pagenumbering{arabic}


\rm
\section{Introduction}

Notably with the advent of the 'Dirichlet brane', which is nowadays
referred to as the D-brane, discovered by Polchinski \cite{Polchinski},
the D-brane theory in various world-volume dimensions has been
widely recognized as an essential ingredient in which to discuss
a variety of non-perturbative aspects of superstring theory
and M-theory, especially the ones concerning the intricate web
of string dualities \cite{Polchinski2}. It is also remakable
that D-branes not only explain the origin of a black hole entropy
\cite{Strominger, Horowitz} but also provide the elementary building
blocks of matrix models \cite{BFSS, IKKT}. Thus it is desirable to
achieve as thorough an understanding of physical properties of
D-branes as possible.

In a previous paper \cite{KO} we have shown that the supersymmetric
and $\kappa$-symmetric Dp-brane actions \cite{Cederwall1, Cederwall2,
Bergshoeff} in a type II supergravity background have the same duality
transformation properties as those in a flat Minkowski background
\cite{Aganagic2}. Specially, it is shown that super D-string transforms
in a covariant way while super D3-brane is self-dual under
the $SL(2,Z)$ duality. Also, D2-brane and D4-brane transform
in ways expected from the relation between type IIA superstring
theory and M-theory.

However, in the study \cite{KO}
we have confined ourselves to a restricted background geometry
where the dilaton and the axion are set to  zero or a constant
although the other fields are considered to be general superfields.
This restriction is obviously unsatisfactory from the following
three reasons.

First, the dilaton and the axion are ordinary interacting particles
in the low energy theory of superstring and thus should be treated
on same footing as the other low energy fields such as
the graviton, the antisymmetric second rank tensor and fermoins.

Second, in a type IIB superstring or supergravity it is well known
that the dilaton and the axion as well as the two 2-form antisymmetric
tensor fields, those are, the NS-NS 2-form and the R-R 2-form,
are doublets of the $SL(2,R)$ Mobius group whereas the graviton
and the 4-form gauge field are singlets in the Einstein metric.
Thus, if we wish to
show that they transform covariantly under the $SL(2,R)$
transformation these local fields must be treated as not constants but
fields. In this context, notice that the vanishing or constant
dilaton and axion imply the vanishing 3-form field strengths,
so in order to understand the transformation rules under the
$SL(2,R)$ we are led to consider the non-constant dilaton and axion.
\footnote{One of authors (I. O.) would like to thank M. Tonin for
pointing out this issue.}

Finally, a study of general dilaton and axion may shed light on
F-theory \cite{F}, a conjectured 12-dimensional quantum field
theory underlying a type IIB superstring.
F-theory is expected to give a systematic description of the
IIB superstring with $\it{non-trivial}$ dilaton and axion
background to which the complex structure of a two-dimensional
torus would correspond.
According to these reasons, the extension of our previous
work \cite{KO} to the non-constant dilaton and axion system
is not only valuable but also quite non-trivial.

In this paper, we will remove our previous restriction on the
dilaton and the axion completely and prove
that various duality transformations found in a type II supergravity
background with the constant or vanishing dilaton and axion
\cite{KO} can be extended to a more general situation where the
dilaton and the axion are superfields depending on the supercoordinates
$Z^M = (X^m, \theta^\mu)$. Therefore, the present study offers a
complete proof that various duality
relations in the super D-brane actions, which were constructed in
\cite{Cederwall1, Cederwall2, Bergshoeff}, exist in
the general on-shell type II supergravity background geometry.

Here let us summarize the results that will be obtained in this
paper.
As in the cases of a flat Minkowski \cite{Aganagic2} and a type II
supergravity backgrounds \cite{KO} with the constant dilaton and
axion, we can show that the super D-string action is the $(m,1)$
string action with the NS-NS charge $m$ and the R-R charge 1.
We prove that the super D3-brane action in the general
background is self-dual under the $SL(2,Z)$ duality
transformation in both the classical and the
quantum-mechanical exact approaches.
Furthermore, it is shown that under a duality transformation
the super D2-brane and D4-brane actions transform in
ways expected from the relation between type IIA superstring
theory and M-theory. Namely, the dual action of the super
D2-brane action is identified as the M2-brane action with
a circular 11th dimension. As for D4-brane, the
double-dimensional reduction of the M5-brane action gives
rise to the dual action of the super D4-brane action.

It is worthwhile to point out the technical differences
between the case of the constant (or zero) dilaton and axion
and that of the non-constant dilaton and axion.
Moving the former on to the latter,
we would encounter at least two complications.
The one is that we have to introduce a new superfield $\Lambda$,
i.e., the dilatino superfield, associated with the dilaton
superfield in a theory, which
gives rise to rather complicated constraint structures
\cite{Cederwall1, Cederwall2}. Indeed, due to this complication
we have confined our consideration to the constant or zero dilaton
and axion in our previous work \cite{KO}.
The second complication that arises is that the non-constant
axion makes it impossible to
use a convenient technique; in the case of the constant axion
we are free to add a theta term to the dual action at the end of
calculations by hand to derive the $SL(2,Z)$ covariance of the
super D-string tension and the $SL(2,Z)$ self-duality of the super
D3-brane \cite{Aganagic2} while in the case of the non-constant
axion we have to deal with the corresponding Wess-Zumino term
from the beginning.
As mentioned in the above, this convenient prescription is far
from complete from the point of view of proof of the $SL(2,R)$
symmetry.

At this point let us comment briefly several previous articles
relating to the study at hand. The duality transformations
of the bosonic D-brane action have been investigated in
Refs.\cite{Schmid, Tseytlin, de Alwis, Lozano}. Afterwards,
supersymmetric D-brane actions with local kappa symmetry
were constructed in a flat background \cite{Aganagic1}
and in a type II supergravity background \cite{Cederwall1,
Cederwall2, Bergshoeff}. Various duality transformations
in a flat background have been clarified in the context of the
super D-brane actions in Ref.\cite{Aganagic2}.
(See also an important paper \cite{Townsend}.)
Recently, motivated by AdS/CFT correspondence \cite{Maldacena}
the duality transformations of super D-string action \cite{Oda1}
and super D3-brane action \cite{Oda2, Kimura, Park} were clarified
in the $AdS_5 \times S^5$ background.
More recently, we have generalized these works to a type II
supergravity background \cite{Oda0, KO}. However, these
works are not completely general in that we have limited
our consideration to the constant or zero dilaton and
axion except super D-string \cite{Oda0}. The main motivation
in this paper is to present a full detail of the duality
transformations of super D-brane actions constructed in
Refs.\cite{Cederwall1, Cederwall2, Bergshoeff} in a general
type II supergravity background.
Incidentally, there are several
interesting papers which attempt to construct actions of
super D-branes with the manifest $SL(2,Z)$-duality and
a dynamical tension \cite{TBC, CW}.
It would be of interest to ask possible
relations between these works and the present study in future.

This article is organized as follows.
Section 2 reviews super D-brane actions in a
general II supergravity background
\cite{Cederwall1, Cederwall2, Bergshoeff}.
In Section 3 it is then proved in both classically and
quantum-mechanically exact manner that the super D-string action
in a type IIB on-shell supergravity background is transformed
to the type IIB Green-Schwarz superstring action \cite{GS} 
in a background with the NS-NS charge $-\lambda$ and the
RR-charge 1.
Section 4 deals with the super D2-brane in a type IIA on-shell
supergravity background and presents that the super D2-brane action
can be transformed to the super M2-brane action with a circular
eleventh dimension by a duality transformation.
In Section 5 we show that the super D3-brane action
in a type IIB on-shell supergravity background is mapped into
itself by an S-duality transformation,
thereby verifying the $SL(2,Z)$ self-duality of the action.
We shall offer both classical and quantum-mechanical proofs
here. In Section 6 it is shown that the super D4-brane
action  becomes identical to the supersymmetric action
which is obtained in terms of double-dimensional reduction
of the super M5-brane action in the eleven dimensional
space-time through a duality transformation.
The final section will be devoted to discussions. In
Appendix a brief review of the $SL(2,R)/SO(2)$ coset description
of the dilaton and the axion in the type IIB supergravity
is given and the type IIB supergravity constraint equations are
shown to be invariant under the $SL(2,R)$ duality transformation.

\section{Super D-brane actions}

Before describing our results for various duality transformations
of super D-brane actions \cite{Cederwall1, Cederwall2, Bergshoeff}
in a general type II supergravity background, we shall
review the salient points of the superspace formulation of the
super Dp-brane theories.
It is well known that super Dp-brane actions are divided
into two pieces, namely, the Dirac-Born-Infeld action and
the Wess-Zumino action in (p+1)-dimensional world-volume.
The former includes the NS-NS two-form,
dilaton and world-volume metric in addition to Abelian gauge
field while the latter action contains the coupling of the
D-brane to the R-R fields. The existence of this Abelian gauge
field is a peculiar feature of D-brane theories.
The two terms are separately invariant under type II superspace
reparametrizations as well as $(p+1)$-dimensional general
coordinate transformations. However, local $\kappa$ symmetry
is achieved by a suitable conspiracy between the two pieces.

Then, the super Dp-brane actions in a general type II on-shell
supergravity background which we consider in this paper
are given by
\begin{eqnarray}
S = S_{DBI} + S_{WZ},
\label{2.1}
\end{eqnarray}
with
\begin{eqnarray}
&{}& S_{DBI} = -  \int_{M_{p+1}} d^{p+1} \sigma \
e^{\frac{p-3}{4} \phi} {\sqrt{- \det ( G_{ij} +
e^{-\frac{1}{2} \phi} {\cal F}_{ij} )}}, \nn\\
&{}& S_{WZ} =  \int_{M_{p+1}} e^{\cal{F}}
\wedge C = \int_{M_{p+1}} \Omega_{p+1}
= \int_{M_{p+2}} I_{p+2},
\label{2.2}
\end{eqnarray}
where $\sigma^i \ (i = 0, 1, \ldots, p)$ are the world-volume
coordinates, $\phi$ the dilaton superfield, and $G_{ij}$ the induced
Einstein metric of the world-volume.
We have defined various quantities as follows:
\begin{eqnarray}
{\cal F} &=& F - b_2, \nn\\
F &=& dA, \nn\\
C &=& \displaystyle \bigoplus_{n=0}^{9} C_{(n)}, \nn\\
I_{p+2} &=& d \Omega_{p+1} = d ( e^{\cal{F}} \wedge C ), \nn\\
M_{p+1} &=& \partial M_{p+2},
\label{2.3}
\end{eqnarray}
where $F$ is the Maxwell field strength 2-form, and the 2-form
$b_2$ is introduced in ${\cal F}$ such that ${\cal{F}}$ is
invariant under
supersymmetry. And the R-R $n$-form fields $C_{(n)}$ are collected
in $C$ with $n$ taking odd integers for type IIA and even integers
for type IIB. The integration of the integrand (\ref{2.2}) involves
forms of various rank; the integral picks out precisely the terms
that are proportional to the volume form of the p-brane world-volume.

In addition, in order to describe the curved target superspace geometry
we have to introduce the superspace vielbein 1-form $E^A$ defined by
\begin{eqnarray}
E^A = dZ^M E_M \ ^A,
\label{2.4}
\end{eqnarray}
with $dZ^M$ denoting the superspace differential $(dX^m, d\theta^\mu)$,
and the torsion 2-form $T^A = DE^A$ as well as the curvature
2-form defined in terms of the spin connection $\omega_A \ ^B$ as
\begin{eqnarray}
R_A \ ^B = d\omega_A \ ^B + \omega_A \ ^C \wedge \omega_C \ ^B.
\label{2.5}
\end{eqnarray}
Note that we have also defined as $M = (m, \mu)$ in curved superspace
while $A = (a, \alpha)$ in flat superspace as usual.
The superspace vielbein 1-form $E^A$ in the tangent space decomposes
under the action of the Lorentz group into a vector $E^a$ and a
spinor $E^{\alpha}$. We shall follow the conventions that the
Lorentz spinor $E^{\alpha}$ is a 32-component Majorana spinor
for the type IIA superspace, on the other hand, a pair of
16-component Majorana-Weyl spinors for the type IIB superspace so
that the latter may be written as $E^{I \alpha}$  with $I$ being the
$N=2$ index ($I=1, 2$).

Then the world-volume metric
$G_{ij}$ is represented by
\begin{eqnarray}
G_{ij} = E_i \ ^a E_j \ ^b \eta_{ab},
\label{2.6}
\end{eqnarray}
where $E_i \ ^A = \partial_i Z^M E_M \ ^A$ and $\eta_{ab}$
= diag$(-,+, \ldots, +)$.

Throughout this paper we use following conventions for superspace
forms. Firstly, a general $n$-form superfield $\Omega_{(n)}$
is expanded as
\begin{eqnarray}
\Omega_{(n)} &=& \frac{1}{n!} dZ^{M_n} \wedge \ldots \wedge dZ^{M_1}
\Omega_{M_1 \ldots M_n}, \nn\\
&=& \frac{1}{n!} E^{A_n} \wedge \ldots \wedge E^{A_1}
\Omega_{A_1 \ldots A_n},
\label{2.7}
\end{eqnarray}
where the superspace differential $dZ^M$ and the superspace
vielbein $E^A$ are antisymmetric with respect to bosonic
coordinates while they are symmetric with respect to fermionic
coordinates.
Secondly, we define the exterior derivative as an operator acting
from the right
\begin{eqnarray}
d (\Omega_{(m)} \wedge \Omega_{(n)}) = \Omega_{(m)} \wedge d\Omega_{(n)}
+ (-)^n d\Omega_{(m)} \wedge \Omega_{(n)}.
\label{2.8}
\end{eqnarray}

Now, following the paper \cite {Cederwall2}, let us define the NS-NS
3-form superfield $H_3$ and the R-R $n$-form superfield
$R$ as \footnote{See the ref.\cite{Bergshoeff}
for type IIA massive supergravity, i.e., $R_{0}=m$}
\begin{eqnarray}
H_{3} &=& db_2, \nn\\
R &=& e^{b_2} \wedge d( e^{-b_2} \wedge C )
= \displaystyle \bigoplus_{n=1}^{10} R_{(n)}.
\label{2.9}
\end{eqnarray}
It is obvious that from these definitions the field strengths
obey the following Bianchi identities
\begin{eqnarray}
dH_{3} &=& 0, \nn\\
e^{b_2} \wedge d( e^{-b_2} \wedge R )
= dR - R \wedge H_{3} &=& 0.
\label{2.10}
\end{eqnarray}

In order to reduce the enormous unconstrained field content
included in the superfields to the field content of the on-shell
type II supergravity theory, one has to impose the constraints
on the torsion and the field strengths by hand, which make various
Bianchi identities coincide with the equations of motion of
type II supergravity.
The nontrivial constraints imposed on the
torsion and field strength components \cite{Cederwall2} take the
following forms for
type IIA:
\begin{eqnarray}
 T_{\alpha\beta} \ ^c &=& 2i \gamma_{\alpha\beta} \ ^c, \nn\\
T_{\alpha\beta} \ ^{\gamma} &=& \frac{3}{2} \delta_{(\alpha}
\ ^{\gamma} \Lambda_{\beta)} + 2 (\gamma_{11})_{(\alpha} \ ^{\gamma}
(\gamma_{11} \Lambda)_{\beta)} - \frac{1}{2} (\gamma_a)_{\alpha\beta}
(\gamma^a \Lambda)^{\gamma}             \nn\\
&{}& + (\gamma_a \gamma_{11})_{\alpha\beta} (\gamma^a \gamma_{11}
\Lambda)^{\gamma} + \frac{1}{4} (\gamma_{ab})_{(\alpha} \ ^{\gamma}
(\gamma^{ab} \Lambda)_{\beta)}, \nn\\
H_{a \alpha\beta} &=& - 2i e^{\frac{1}{2} \phi} (\gamma_{11}
\gamma_a)_{\alpha\beta}, \nn\\
H_{ab \alpha} &=& e^{\frac{1}{2} \phi} (\gamma_{ab} \gamma_{11}
\Lambda)_{\alpha}, \nn\\
R_{(n) a_1 \ldots a_{n-2} \alpha\beta} &=&  2i e^{\frac{n-5}{4} \phi}
(\gamma_{a_1 \ldots a_{n-2}} (\gamma_{11})^{\frac{n}{2}})_{\alpha\beta},
\nn\\
R_{(n) a_1 \ldots a_{n-1} \alpha} &=& - \frac{n-5}{2}
e^{\frac{n-5}{4} \phi} (\gamma_{a_1 \ldots a_{n-1}}
(-\gamma_{11})^{\frac{n}{2}} \Lambda)_{\alpha},
\label{2.11}
\end{eqnarray}
and for IIB:
\begin{eqnarray}
T_{\alpha\beta} \ ^c &=& 2i \gamma_{\alpha\beta} \ ^c, \nn\\
T_{\alpha\beta} \ ^{\gamma} &=& - ({\cal I})_{(\alpha}
\ ^{\gamma} ({\cal I} \Lambda)_{\beta)} + ({\cal K})_{(\alpha}
\ ^{\gamma} ({\cal K} \Lambda)_{\beta)}  \nn\\
&{}& + \frac{1}{2} (\gamma_a {\cal I})_{\alpha\beta} (\gamma^a
{\cal I} \Lambda)^{\gamma} - \frac{1}{2} (\gamma_a {\cal K})_{\alpha\beta}
(\gamma^a {\cal K} \Lambda)^{\gamma}, \nn\\
H_{a \alpha\beta} &=& - 2i e^{\frac{1}{2} \phi}
({\cal K} \gamma_a)_{\alpha\beta}, \nn\\
H_{ab \alpha} &=& e^{\frac{1}{2} \phi} (\gamma_{ab} {\cal K}
\Lambda)_{\alpha}, \nn\\
R_{(n) a_1 \ldots a_{n-2} \alpha\beta} &=&  2i e^{\frac{n-5}{4} \phi}
(\gamma_{a_1 \ldots a_{n-2}} {\cal K}^{\frac{n-1}{2}}
{\cal E})_{\alpha\beta}, \nn\\
R_{(n) a_1 \ldots a_{n-1} \alpha} &=& - \frac{n-5}{2} e^{\frac{n-5}{4} \phi}
(\gamma_{a_1 \ldots a_{n-1}} {\cal K}^{\frac{n-1}{2}}
{\cal E} \Lambda)_{\alpha},
\label{2.12}
\end{eqnarray}
where the round brackets enclosing indices denote symmetrization
with 'strength one', and then $\Lambda_{\alpha} \equiv \frac{1}{2}
\partial_{\alpha} \phi$, and ${\cal E}$, ${\cal I}$, and ${\cal K}$
describing the $SL(2,R)$ matrices are defined in terms of the
conventional Pauli matrices $\sigma_i$ as follows:
\begin{eqnarray}
{\cal E} = i \sigma_2 = \pmatrix{
0  & 1 \cr -1 & 0 \cr }, \ {\cal I} = \sigma_1 = \pmatrix{
0  & 1 \cr 1 & 0 \cr },  \ {\cal K} = \sigma_3 = \pmatrix{
1  & 0 \cr 0 & -1 \cr }.
\label{2.13}
\end{eqnarray}

Based on this formulation of the super Dp-brane actions in a type
II on-shell supergravity background, we shall explore various
duality symmetries in subsequent sections.

\section{The super D-string}

In this section we would like to consider the super D-string
(i.e. the super D1-brane) first. The super D2, D3 and D4-branes
will be treated in order in subsequent sections. In these sections
we shall prove various duality symmetries of the super D-brane
actions in a type II on-shell supergravity background. The corresponding
proofs have been already done in a flat Minkowskian background
 \cite{Aganagic2} and a type II supergravity background
with the constant or zero dilaton and axion \cite{KO}.
A partial result was also presented in a type IIB
supergravity background
with the non-constant dilaton and axion \cite{Oda0}.
The purpose of this section is to offer the full detail
and address the issue of the $SL(2,Z)$ symmetry in such
a most general background.
Our study appears to be quite important for future
development in string theory and M-theory since the global
discrete symmetries such as the $SL(2,Z)$ S-duality are nowadays
believed to be exact symmetries in still mysterious underlying theory
\cite{Hull, Witten2} so that these symmetries should be valid
even in a curved background geometry.

In the case at hand, by making use of the superspace convention
(\ref{2.7}) the action (\ref{2.1}), (\ref{2.2}) and the
relevant constraints (\ref{2.12}) for type IIB reduce to
\begin{eqnarray}
S &=& S_{DBI} + S_{WZ}, \nn\\
S_{DBI} &=& - \int_{M_2} d^2 \sigma e^{-\frac{1}{2} \phi}
\sqrt{- \det ( G_{ij} + e^{-\frac{1}{2} \phi}
{\cal F}_{ij} )}, \nn\\
S_{WZ} &=& \int_{M_2 = \partial M_3} (C_2 + C_0 {\cal F})
= \int_{M_3} I_3,
\label{3.1}
\end{eqnarray}
\begin{eqnarray}
H_{3} &=& db_2 = i e^{\frac{1}{2} \phi} \bar{E} \wedge \hat{E}
\wedge {\cal K} E + \frac{1}{2} e^{\frac{1}{2} \phi}
\bar{E} \wedge \gamma_{ab} {\cal K} \Lambda E^b \wedge E^a , \nn\\
R_{(1)} &=& dC_0 =  2 e^{-\phi} \bar{E} {\cal E} \Lambda, \nn\\
R_{(3)} &=& dC_2 - H_3 C_0 \nn\\
&=& - i e^{-\frac{1}{2} \phi} \bar{E} \wedge \hat{E} \wedge
{\cal I} E + \frac{1}{2} e^{-\frac{1}{2} \phi} \bar{E} \wedge
\gamma_{ab} {\cal I} \Lambda E^b \wedge E^a,
\label{3.d}
\end{eqnarray}
where $\bar{E}$, $\hat{E}$ and $E$
represent the Dirac conjugate of $E^{I\alpha}$, $E^a \gamma_a$
and $E^{I\alpha}$, respectively.
The equations in (\ref{3.1}) and (\ref{3.d}) will serve as our
basis for all subsequent considerations in this section.

\subsection{The classical analysis}

In this subsection, we turn our attention to a duality
transformation of the super D-string action in a classical
approach. Next subsection will provide a quantum-mechanical
exact treatment of the same model.

Having exhibited the explicit forms of the classical action
and the constraints of the super D-string, the next task
is to construct a dual action out of them. The strategy
for this purpose is now standard, but for definiteness let
us present the detailed exposition.
As a first step, one needs to incorporate a Lagrange multiplier
field $\tilde{H}^{ij} = - \tilde{H}^{ji}$ into the classical
action as follows \cite{Tseytlin}:
\begin{eqnarray}
S &=& - \int_{M_2} d^2 \sigma e^{-\frac{1}{2} \phi}
\sqrt{- \det ( G_{ij} + e^{-\frac{1}{2} \phi}
{\cal F}_{ij} )} + \int_{M_2} (C_2 + C_0 {\cal F}) \nn\\
&{}& + \int_{M_2} d^2 \sigma \frac{1}{2} \tilde{H}^{ij}
(F_{ij} - 2 \partial_i A_j),
\label{3.2}
\end{eqnarray}
and then regard $F_{ij}$ as an independent superfield.
The equation of motion of $A_i$ gives rise to $\partial_i
\tilde{H}^{ij} = 0$ whose classical solution is found to be
$\tilde{H}^{ij} =
\varepsilon^{ij} \lambda$ with a constant scalar superfield
$\lambda$.

Next, substituting this classical solution into the original
action (\ref{3.2}) gives us an action $S' = S_1 + S_{2D}$, where
\begin{eqnarray}
S_1 &=& \int_{M_2} d^2 \sigma \left[ - e^{-\frac{1}{2} \phi}
\sqrt{- \det ( G_{ij} + e^{-\frac{1}{2} \phi}
{\cal F}_{ij} )} + \frac{1}{2} (\lambda + C_0)
\varepsilon^{ij} {\cal F}_{ij} \right], \nn\\
S_{2D} &=& \int_{M_2} (C_2 + \lambda b_2).
\label{3.3}
\end{eqnarray}
The important procedure to get a dual action is to solve the
equation of motion obtained by varying $F_{ij}$
to express the action $S'$ in terms of
$\lambda$ instead of $F_{ij}$. Note that since $S_{2D}$
does not include $F_{ij}$ in it, this part of the action is
manifestly unaffected by the duality transformation.
A simple calculation leads to
\begin{eqnarray}
{\cal F}_{01} = - e^{\frac{1}{2} \phi}
\frac{\lambda + C_0}{\sqrt{e^{-2 \phi}+(\lambda + C_0)^2}}
\sqrt{- \det{G_{ij}}},
\label{3.4}
\end{eqnarray}
where we have used the formula holding for $2 \times 2$ matrices
\begin{eqnarray}
\det ( G_{ij} + e^{-\frac{1}{2} \phi} {\cal F}_{ij} )
&=& \det{G_{ij}} + e^{- \phi} {\cal F}_{01}^2 \nn\\
&=& \det{G_{ij}} + \frac{1}{4} e^{- \phi} (\varepsilon^{ij}
{\cal F}_{ij})^2.
\label{3.5}
\end{eqnarray}
It follows that inserting Eq.(\ref{3.4}) to $S_1$ together with $S_{2D}$
yields the dual action of the super D-string action
\begin{eqnarray}
S_D = - \int_{M_2} d^2 \sigma e^{\frac{1}{2} \phi}
\sqrt{e^{-2 \phi} + (\lambda + C_0)^2}
\sqrt{- \det G_{ij}} + \int_{M_2} (C_2 + \lambda b_2).
\label{3.6}
\end{eqnarray}

We may interpret the dual action (\ref{3.6}) in the following way.
If we regard the Wess-Zumino term in (\ref{3.6}),
$\int_{M_2} (C_2 + \lambda b_2)$, as the source term
for the NS-NS  and the R-R 2-form potentials, it turns out that
this dual action carries the NS-NS charge $-\lambda$ and
the R-R charge 1 \footnote{Here we have used the fact
that $(b_2, -C_2)$ is the $SL(2,R)$ doublet as shown around
the end of this section.}.
Then provided that we identify the constant
scalar superfield $- \lambda = m$ with the NS-NS charge corresponding
to the $(m,1)$ string \footnote{See the next subsection about the
reason why the scalar $\lambda$ becomes an integer $m$.},
the result obtained above, $\sqrt{e^{-2 \phi}
+ (m - C_0)^2}$, agrees with the tension formula for the
$SL(2,Z)$ S-duality spectrum of strings in the type IIB superstring
\cite{Schwarz}
\footnote{For comparison of the string tension,
we need to consider the string metric defined as $G_{ij}^{(s)}
= e^{\frac{1}{2} \phi} G_{ij}$ where $G_{ij}$ is the Einstein
metric.}.
This identification means that the D-string action may be actually
the action for an arbitrary number of 'fundamental' IIB strings
bound to a single D-string. This singleness of D-string of
course reflects the $U(1)$ character of the Abelian gauge
field since a system of $N$ coincident Dp-branes would be
described by the non-abelian version of
the Dirac-Born-Infeld action \cite{DBI}.

 Now let us examine the implication of Eq.(\ref{3.6}) in more detail
and consider the exact connection with the type IIB Green-Schwarz 
superstring action.
For that purpose we rewrite the Wess-Zumino term in (\ref{3.6}) in
the form
\begin{eqnarray}
d\Omega_D &\equiv& d (C_2 + \lambda b_2) \nn\\
&=& R_{(3)} + (\lambda + C_0) H_3 \nn\\
&=& i \bar{E} \wedge \gamma_a \left[e^{-\frac{1}{2} \phi} {\cal I}
- (\lambda + C_0) e^{\frac{1}{2} \phi} {\cal K} \right]
E \wedge E^a  \nn\\
&{}& + \frac{1}{2} \bar{E} \wedge \gamma_{ab} \left[e^{-\frac{1}{2}
\phi}{\cal I} +  (\lambda + C_0) e^{\frac{1}{2} \phi} {\cal K}
\right] \Lambda E^b \wedge E^a,
\label{3.e}
\end{eqnarray}
where Eq.(\ref{3.d}) was used. Although the two matrices in
the square brackets which are $2 \times 2$ hermitian matrices,
have the same eigenvalues $\pm e^{-\frac{1}{2} \phi}
\sqrt{1 + (\lambda + C_0)^2 e^{2 \phi}}$, these matrices are not
mutually commutable so that they cannot be diagonalized
simultaneously.

In the case of constant dilaton and axion the last term in
(\ref{3.e}) vanishes and the first term can be diagonalized by
a suitable $SO(2)$ spinor-rotation. As a result, (\ref{3.e})
can be written as
\begin{eqnarray}
d\Omega_{D} =
e^{\frac{1}{2} \phi}
\sqrt{e^{-2 \phi} + (\lambda + C_0)^2}
i\bar{E}^{\prime}\wedge
\gamma_{a}E^{a}\wedge{\cal K}E^{\prime}.
\nonumber
\end{eqnarray}
{}From this form of the dual action, it has been concluded previously
\cite{Aganagic2,Oda0,KO} that the dual action of D-string is
equivalent to the type IIB Green-Schwarz superstring with the
$SL(2,Z)$ covariant string tension $\sqrt{e^{-2 \phi}
+ (\lambda + C_0)^2}$ \cite{Schwarz}.
But this argument is a little naive because in this case
we can change the string tension at will through the field
redefinitions.

In the present formulation, however, we have to take notice of at
least two problems in this argument.
Firstly in the process of the diagonalization the $SO(2)$
spinor-rotation must be accompanied by some associated $SL(2,R)$
transformation of the background fields
in order to preserve the type IIB supergravity constraint
equations; namely type IIB supergravity equations of motion.
Secondly since in the general case of non-constant dilaton and axion
the last term in (\ref{3.e}) does not vanish,
two terms in (\ref{3.e}) can not be simultaneously diagonalized
by some $SO(2)$ spinor-rotation.

It is remakable to see that these problems are closely related and
resolved in the following way.
To do so let us rewrite (\ref{3.6}) in the form
\begin{eqnarray}
S_D^{\prime} = - \int_{M_2} d^2 \sigma e^{\frac{1}{2}\phi^{\prime}}
\sqrt{- \det G_{ij}} + \int_{M_2} b_{2}^{\prime},
\label{3.a}
\end{eqnarray}
where
\begin{eqnarray}
e^{\frac{1}{2}\phi^{\prime}}
&=&e^{\frac{1}{2} \phi}
\sqrt{e^{-2\phi}+ (\lambda + C_0)^2}, \nonumber \\
b_{2}^{\prime}
&=& C_2 + \lambda b_2.
\label{3.b}
\end{eqnarray}
According to the $SL(2,R)/SO(2)$ coset description of dilaton and
axion in type IIB supergravity reviewed in Appendix, Eq.(\ref{3.b})
is just the $SL(2,Z)$ transformation $\tau\rightarrow
\tau^{\prime}=\frac{-1}{\tau + \lambda}$
associated with the $SL(2,Z)$ matrix
\begin{equation}
S=
\left(
\begin{array}{cc}
0 & -1 \\
1 & \lambda
\end{array}
\right), \
(S^{T})^{-1}=
\left(
\begin{array}{cc}
\lambda & -1 \\
1 & 0
\end{array}
\right),
\label{3.c}
\end{equation}
where $(b_{2},-C_{2})$ belongs to an $SL(2,R)$ doublet which
transforms like $B$ in (\ref{A-11}).

As discussed in Appendix the type IIB supergravity constraints
are invariant under the $SL(2,R)$ transformation combined with
some suitable $SO(2)$ spinor-rotation and therefore
$b_{2}^{\prime}$ satisfies the type IIB supergravity constraint
equation for the NS-NS 2-form.
To compare with the Wess-Zumino term of the type IIB Green-Schwarz
superstring action, let us take an exterior derivative of the
integrand in the Wess-Zumino term, which is calculated as follows:
\begin{eqnarray}
d \Omega_D^{\prime} &\equiv&  db_2^{\prime} \nn\\
&=&e^{\frac{1}{2}\phi^{\prime}}
\left[i\bar{E}^{\prime} \wedge \gamma_{a}E^{a}
\wedge{\cal K}E^{\prime}
+ \frac{1}{2} \bar{E}^{\prime} \wedge \gamma_{ab}{\cal K}
\Lambda^{\prime} E^b \wedge E^a \right].
\label{3.7}
\end{eqnarray}
where $E^{\prime}$ and $\Lambda^{\prime}$ are the $SO(2)$
spinor-rotated supervielbein and dilatino superfield, respectively.
The spinor-rotation angle is given by (\ref{A-17}) and (\ref{3.c}).
Thus it follows that the dual action (\ref{3.a}) is the type IIB
Green-Schwarz superstring action with the unit string tension,
$T=1$ in the $SL(2,R)$ transformed type IIB supergravity
background \cite{Grisaru}.
(Note that this $SL(2,Z)$ transformed background
has the NS-NS charge $-\lambda$ and the RR-charge 1.)
In this way the dual action (\ref{3.6}) is
precisely transformed into the type IIB Green-Schwarz superstring
action by a $SL(2,Z)$ duality transformation (combined with a
suitable $SO(2)$ spinor-rotation).

Of course, this does not imply that the dual action (\ref{3.6})
of the super D-string is equivalent to the Green-Schwarz
superstring action with the unit NS-NS charge, for
the $SL(2,Z)$ transformation is not a symmetry in
the case of string theory, as opposed to D3-brane case.
Instead, the above demonstration means that the dual action
(\ref{3.6}) is indeed  nothing but the type IIB superstring
action in a background with the NS-NS charge $-\lambda$ and
the RR-charge 1.

As alluded in Section 1, in order to obtain this result,
we have not adopted the convenient technique for which
the Wess-Zumino action containing the constant axion, which is a
topological theta term, is taken into account at the end of
calculations in order to produce the $SL(2,Z)$ covariant tension,
instead treated its non-trivial Wess-Zumino action due to the
non-constant character of the axion field from the beginning
in a direct manner.

\subsection{The quantum analysis}

The results we have obtained in the previous subsection
unfortunately rely on the classical analysis
in the sense that we have made use
of the field equations at least in the two stages, namely,
we have used the field equations of $A_i$ and $F_{ij}$.
Now we are ready to present a quantum-mechanical exact proof
of $SL(2,Z)$ S-duality of the super D-string action
in a general ten dimensional IIB supergravity background.
(The results in this subsection were briefly reported in a
short article \cite{Oda0}.)

To this aim, let us utilize the first-order Hamiltonian form
of path integral following the techniques developed in
\cite{de Alwis, Oda4}.
As a first step of the Hamiltonian formalism, let us introduce
the canonical conjugate momenta $\pi^i$ corresponding to the
gauge field $A_i$ defined as
\begin{eqnarray}
\pi^i \equiv \frac{\partial S}
{\partial \dot{A}_i},
\label{3.9}
\end{eqnarray}
where the dot denotes the time derivative.
Then the canonical conjugate momenta $\pi^i$ are calculated to be
\begin{eqnarray}
\pi^0 = 0, \ \pi^1 = e^{-\frac{3}{2} \phi}
\frac{{\cal F}_{01}}
{\sqrt{-\det ( G_{ij} + e^{-\frac{1}{2} \phi} {\cal F}_{ij} )}}
+ C_0,
\label{3.10}
\end{eqnarray}
where the former equation implies the existence of the $U(1)$
gauge invariance.  From the latter equation $\dot{A}_1$ can be
expressed in terms of $\pi^1$ whose result is given by
\begin{eqnarray}
\dot{A}_1 = e^{\frac{1}{2} \phi}
\frac{\pi^1 - C_0}{\sqrt{(\pi^1 - C_0)^2 + e^{-2 \phi}}}
\sqrt{- \det{G_{ij}}} + b_{01} + \partial_1 A_0.
\label{3.11}
\end{eqnarray}
Then we will see that the Hamiltonian density takes the form
\begin{eqnarray}
{\cal H} = e^{\frac{1}{2} \phi} \sqrt{e^{-2 \phi}+(\pi^1 - C_0)^2}
\sqrt{- \det G_{ij}} - A_0 \partial_1 \pi^1
+ \partial_1 (A_0 \pi^1) + \pi^1 b_{01} - C_{01}.
\label{3.12}
\end{eqnarray}

Now the partition function is defined by the first-order
Hamiltonian form with respect to only the gauge field as follows:
\begin{eqnarray}
Z &=& \frac{1}{\int {\cal D}\pi^0} \int {\cal D}\pi^0
{\cal D}\pi^1 {\cal D}A_0 {\cal D}A_1
\exp{ i \int d^2 \sigma ( \pi^1 \partial_0 A_1 - {\cal H} ) } \nn\\
&=& \int {\cal D}\pi^1 {\cal D}A_0 {\cal D}A_1
\exp{ i \int d^2 \sigma} \nn\\
&{}& \times \left[ - A_1 \partial_0 \pi^1
+ A_0 \partial_1 \pi^1
-  e^{\frac{1}{2} \phi} \sqrt{e^{-2 \phi}+ (\pi^1 - C_0)^2}
\sqrt{- \det G_{ij}}
- \pi^1 b_{01} + C_{01} \right],
\label{3.13}
\end{eqnarray}
where we have taken the boundary condition for $A_0$ such that
the surface term identically vanishes.
Then we can carry out the integrations over $A_i$ explicitly,
which gives rise to $\delta$ functions
\begin{eqnarray}
Z &=& \int {\cal D}\pi^1 \delta(\partial_0 \pi^1)
\delta(\partial_1 \pi^1) \exp{ i \int d^2 \sigma} \nn\\
&{}& \times \left[ -e^{\frac{1}{2} \phi}
\sqrt{e^{-2\phi}+(\pi^1 - C_0)^2}\sqrt{- \det G_{ij}}
+ C_{01} - \pi^1 b_{01} \right].
\label{3.14}
\end{eqnarray}
The existence of the $\delta$ functions reduces the integral
over $\pi^1$ to the one over only its zero-modes. If we require
that one space component is compactified on a circle, these
zero-modes are quantized to be integers \cite{Witten}.
As a consequence, the partition function becomes
\begin{eqnarray}
Z &=& \displaystyle{ \sum_{m \in {\bf Z}} } \exp{ i \int d^2 \sigma
\left[ - e^{\frac{1}{2} \phi} \sqrt{-e^{2 \phi}+ (m - C_0)^2}
\sqrt{- \det G_{ij}} + C_{01} - m b_{01} \right] },
\label{3.15}
\end{eqnarray}
from which we can read off the effective action
\begin{eqnarray}
S = \int d^2 \sigma \left( -e^{\frac{1}{2} \phi}
\sqrt{e^{-2 \phi} + (m - C_0)^2} \sqrt{- \det
G_{ij}} + C_{01} - m b_{01} \right).
\label{3.16}
\end{eqnarray}
Note that if we replace $m$ with $-\lambda$ this action precisely
becomes equivalent to (\ref{3.6}).
In this way, we have derived the dual action of the super D-string
action in a type IIB on-shell supergravity background geometry in
quantum-mechanical exact manner.

To close this section let us present an interesting observation
which will play an important role especially in proving
the self-duality of the super D3-brane action in Sec.5.
Notice that the original action (\ref{3.1}) possesses the
following two symmetries;
\begin{eqnarray}
C_{0}&\rightarrow& C_{0}^{\prime}=C_{0}+1,
\nonumber \\
b_{2}&\rightarrow& b_{2}^{\prime}=b_{2}, \ C_{2}\rightarrow
C_{2}^{\prime}=C_{2}-{\cal F}=C_{2}+b_{2}-F,
\label{3-17}
\end{eqnarray}
and
\begin{eqnarray}
C_{0}&\rightarrow& C_{0}^{\prime}=C_{0}+1,
\nonumber \\
b_{2}&\rightarrow& b_{2}^{\prime}=b_{2}, \ C_{2}\rightarrow
C_{2}^{\prime}=C_{2}+b_{2},
\label{3-18}
\end{eqnarray}
which stem from the structure of the Wess-Zumino action.
The super D-string action (\ref{3.1}) is exactly invariant
under (\ref{3-17}) and invariant only up to the topological
term F under (\ref{3-18}).
These symmetries do not affect the type IIB supergravity
constraints (\ref{3.d}), which implies that these are real
symmetries of the theory.
Of course, it is obvious that the partition function
(\ref{3.15})
is also invariant under these symmetries.
Noticing $F=dA$ the transformation (\ref{3-17}) may be
interpreted as the transformation (\ref{3-18}) combined
with a gauge transformation of $C_{2}$ whose gauge
parameter is the world-volume Abelian gauge field itself.

Then it is natural to ask what the meaning of these symmetries
is. In the following we shall follow the notations in Appendix.
First, when we introduce the complex variable $\tau$ by
$\tau = C_{0}+ie^{-\phi}$,
the shift of $C_{0}$ corresponds to $\tau \rightarrow \tau +1$,
being an element of $SL(2,R)$ associated with $SL(2,R)$ matrix
\begin{equation}
S=
\left(
\begin{array}{cc}
1 & 1 \\
0 & 1
\end{array}
\right),
(S^{T})^{-1}=
\left(
\begin{array}{cc}
1 & 0\\
-1 & 1
\end{array}
\right),
\label{3-20}
\end{equation}
Next we should also investigate how this symmetry acts on the
other superfields in the theory.
Comparing the transformation (\ref{3-18}) with (\ref{A-11}),
$(b_{2},-C_{2})$ is expected to form an $SL(2,R)$ doublet
which transforms like $B$ in (\ref{A-11}).
This statement is consistent with the result obtained in the
previous subsection and also supported in Sec.5 by the fact
that under the weak-strong duality
$\tau\rightarrow -\frac{1}{\tau}$,
$(b_{2},-C_{2})$ correctly transforms as an $SL(2,R)$ doublet.

It is illuminating that super D-brane actions in the general
type IIB supergravity background has a shift symmetry
$\tau\rightarrow \tau+1$ as expected from the fact that this
is a symmetry of pertubation theory. This fact has been thus far
understood in the target space approach where the R-R scalar
$C_{0}$ appears only through its field strength. But in the
present world-volume approach this comes from the non-trivial
symmetry of the super D-brane action. We wish to stress that
it is possible to prove this interesting observation only
in the present formalism within the context of the world-volume
theory as mentioned in Sec.1.

\section{The super D2-brane}

Next we turn to the $\it{classical}$ derivation of a duality
transformation between the super D2-brane
(i.e., the super D-membrane) in a type IIA
supergravity background and the super M2-brane in eleven dimensional
supergravity. The authors in a paper \cite{Bergshoeff} have already
dealt with this problem from a different viewpoint. The method adopted
there is to start with the super M2-brane in eleven dimensions, achieve
the dimensional reduction to ten dimensions a la Kaluza-Klein ansatz,
then perform a duality transformation for the purpose of getting the
super D2-brane action and its $\kappa$-symmetry. Our method is
similar to that of Aganagic et al. \cite{Aganagic2} where
the above arguments were reversed, namely, the super M2-brane action
was obtained from starting with
the super D2-brane action through a duality transformation
of the world-volume gauge field. The case of the constant or zero
dilaton was already considered in our previous work \cite{KO}.
Making use of the formulation explained in Sec.2, we focus on the
problem of how our previous work should be extended to the case of
the non-constant dilaton superfield.

{}From Eqs.(\ref{2.1}) and (\ref{2.2}), the super D2-brane action is
of the form
\begin{eqnarray}
S &=& S_{DBI} + S_{WZ} + S_{\tilde{H}}, \nn\\
S_{DBI} &=& - \int_{M_3} d^3 \sigma e^{-\frac{1}{4} \phi}
\sqrt{- \det ( G_{ij} + e^{-\frac{1}{2} \phi} {\cal F}_{ij} )}, \nn\\
S_{WZ} &=& \int_{M_3} ( C_3 + C_1 \wedge {\cal F} )
= \int_{M_4} I_4, \nn\\
S_{\tilde{H}} &=& \int_{M_3} d^3 \sigma \frac{1}{2} \tilde{H}^{ij}
( F_{ij} - 2 \partial_i A_j ),
\label{4.1}
\end{eqnarray}
where we have added $S_{\tilde{H}}$ to the original action in
order to perform a duality transformation. Moreover, in this case
the constraints (\ref{2.11}) on the field strengths reduce to
\begin{eqnarray}
H_3 &=& db_2 = i e^{\frac{1}{2} \phi} \bar{E} \wedge \gamma_{11}
\hat{E} \wedge E + \frac{1}{2} e^{\frac{1}{2} \phi} \bar{E} \wedge
\gamma_{ab} \gamma_{11} \Lambda E^b \wedge E^a , \nn\\
R_{(2)} &=& dC_1 = i e^{-\frac{3}{4} \phi} \bar{E} \wedge \gamma_{11} E
- \frac{3}{2} e^{-\frac{3}{4} \phi} \bar{E} \wedge \gamma_a \gamma_{11}
\Lambda E^a, \nn\\
R_{(4)} &=& dC_3 + H_3 \wedge C_1 \nn\\
&=& \frac{i}{2} e^{-\frac{1}{4} \phi} \bar{E} \wedge \gamma_{ab} E
\wedge E^b \wedge E^a + \frac{1}{12} e^{-\frac{1}{4} \phi}
\bar{E} \wedge \gamma_{abc} \Lambda E^c \wedge E^b \wedge E^a,
\label{4.2}
\end{eqnarray}
where (\ref{2.7}), (\ref{2.8}) and (\ref{2.9}) were also utilized.
Of course, $C_1$ and $C_3$ are determined by solving Eq.(\ref{4.2}).

Varying $A_i$ in the action (\ref{4.1}) gives the equation of motion,
$\partial_i \tilde{H}^{ij} = 0$ like the super D-string case.
Then the solution is obviously
given by $\tilde{H}^{ij} = \epsilon^{ijk} \partial_k B$ with $B$
being a scalar superfield.
Next, substituting this classical solution into the original
action (\ref{4.1}) leads to an action $S' = S_1 + S_{2D}$, where
\begin{eqnarray}
S_1 &=& \int_{M_3} d^3 \sigma \left[ - e^{-\frac{1}{4} \phi}
\sqrt{- \det ( G_{ij} + e^{-\frac{1}{2} \phi} {\cal F}_{ij} )}
+ \frac{1}{2} \varepsilon^{ijk} (\partial_k B + C_k)
{\cal F}_{ij} \right], \nn\\
S_{2D} &=& \int_{M_3} (C_3 + b_2 \wedge dB).
\label{4.3}
\end{eqnarray}
Since $S_{2D}$ does not include $F_{ij}$, it is manifestly duality
invariant. Solving the equation of motion for $F_{ij}$ in order to
rewrite the action $S_1$ in terms of $B$ instead of $F_{ij}$,
we arrive at the dual action $S_D$ of (\ref{4.1})
\begin{eqnarray}
S_D = - \int_{M_3} d^3 \sigma e^{-\frac{1}{4} \phi}
\sqrt{- \det G'_{ij}} + \int_{M_3} ( C_3 + b_2 \wedge dB ),
\label{4.4}
\end{eqnarray}
where we have defined as
\begin{eqnarray}
G'_{ij} = G_{ij} + e^{\frac{3}{2} \phi} (\partial_i B + C_i)
(\partial_j B + C_j).
\label{4.5}
\end{eqnarray}
Incidentally, in order to derive the dual action we have used
the mathematical formulas holding for $3 \times 3$ matrices
\begin{eqnarray}
\det ( G_{ij} + A_i A_j ) &=& (\det G_{ij}) \times ( 1 + G^{ij}
A_i A_j ), \nn\\
\det ( G_{ij} + {\cal F}_{ij} ) &=& (\det G_{ij})
\times ( 1 + \frac{1}{2} G^{ij} G^{kl} {\cal F}_{ik} {\cal F}_{jl} ),
\label{4.6}
\end{eqnarray}
where ${\cal F}_{ij} = -{\cal F}_{ji}$.

Eq.(\ref{4.5}) implies the identification $dB + C_1 =
e^{-\frac{3}{4} \phi} E^{11}$,
in other words, identifying the world-volume scalar with the
coordinate of a compact extra target-space dimension.
Consequently, the Dirac-Born-Infeld action in Eq.(\ref{4.4})
takes the standard form for the induced metric of the M2-brane.
The remaining work is to show that the second term in the right
hand side of Eq.(\ref{4.4}) equals to the expression
for the Wess-Zumino term of the super M2-brane. Taking
the exterior derivative and using the relation (\ref{4.2})
we can evaluate this term as follows:
\begin{eqnarray}
d \Omega_D &\equiv& d ( C_3 + b_2 \wedge dB ) \nn\\
&=& R_{(4)} + (dB + C_1) \wedge H_3 \nn\\
&=& \frac{i}{2} e^{-\frac{1}{4} \phi} \bar{E} \wedge \gamma_{ab} E
\wedge E^b \wedge E^a + \frac{1}{12} e^{-\frac{1}{4} \phi} \bar{E}
\wedge \gamma_{abc} \Lambda E^c \wedge E^b \wedge E^a  \nn\\
&{}& + (i e^{-\frac{1}{4} \phi} \bar{E} \wedge \gamma_{11} \gamma_a
E \wedge E^a - \frac{1}{2} e^{-\frac{1}{4} \phi}
\bar{E} \wedge \gamma_{ab} \gamma_{11} \Lambda E^b \wedge E^a)
\wedge E^{11}   \nn\\
&=& \frac{i}{2} e^{-\frac{1}{4} \phi} \bar{E} \wedge
\gamma_{\hat{a}\hat{b}} E \wedge E^{\hat{b}} \wedge E^{\hat{a}}
\nn\\
&{}& + \frac{1}{12} e^{-\frac{1}{4} \phi} \bar{E}
\wedge \gamma_{abc} \Lambda E^c \wedge E^b \wedge E^a
- \frac{1}{2} e^{-\frac{1}{4} \phi}
\bar{E} \wedge \gamma_{ab} \gamma_{11} \Lambda E^b \wedge E^a
\wedge E^{11} \nn\\
&\equiv& d \Omega_1 + d \Omega_2 ,
\label{4.7}
\end{eqnarray}
where $\hat{a} \equiv (a, 11)$ denotes 11 dimensional index,
and $d \Omega_1$ and $d \Omega_2$ are respectively defined
as $d \Omega_1 = \frac{i}{2} e^{-\frac{1}{4} \phi} \bar{E} \wedge
\gamma_{\hat{a}\hat{b}} E \wedge E^{\hat{b}} \wedge E^{\hat{a}}$
and $d \Omega_2 = \frac{1}{12} e^{-\frac{1}{4} \phi} \bar{E}
\wedge \gamma_{abc} \Lambda E^c \wedge E^b \wedge E^a
- \frac{1}{2} e^{-\frac{1}{4} \phi}
\bar{E} \wedge \gamma_{ab} \gamma_{11} \Lambda E^b \wedge E^a
\wedge E^{11}$.
Accordingly, the
dual action (\ref{4.4}) of the super D2-brane can be written as
\begin{eqnarray}
S_D = - \int_{M_3} d^3 \sigma e^{-\frac{1}{4} \phi}
\sqrt{- \det G'_{ij}}
+ \int_{M_3} \Omega_1 + \int_{M_3} \Omega_2,
\label{4.8}
\end{eqnarray}
where $G'_{ij} = E_i^{\hat{a}} E_j^{\hat{b}} \eta_{\hat{a}\hat{b}}$
\footnote{$\eta_{\hat{a}\hat{b}}$ is the 11 dimensional flat metric
defined as diag$(-,+, \ldots,+,+)$.}.
In order to make (\ref{4.8}) coincide with the super
M2-brane action where there is no
dilaton, we should remove the dilaton superfield except $\Omega_2$
by rescaling the other superfields.
It then turns out that the following rescaling makes a job
\begin{eqnarray}
E^{\hat{a}} \rightarrow  e^{\frac{1}{12} \phi} E^{\hat{a}}, \
E^{\alpha} \rightarrow  e^{\frac{1}{24} \phi} E^{\alpha}, \
\Lambda \rightarrow  e^{\frac{1}{24} \phi} \Lambda.
\label{4.9}
\end{eqnarray}
After this rescaling, the dual action takes the form
\begin{eqnarray}
S_D = - \int_{M_3} d^3 \sigma
\sqrt{- \det G^{11}_{ij}} + \int_{M_3} \Omega^{11}
+ \int_{M_3} \Omega',
\label{4.10}
\end{eqnarray}
where $d \Omega^{11} = \frac{i}{2} \bar{E} \wedge
\gamma_{\hat{a}\hat{b}} E \wedge E^{\hat{b}} \wedge E^{\hat{a}}$
and $d \Omega' = \frac{1}{12} e^{\frac{1}{12} \phi} \bar{E}
\wedge \gamma_{abc} \Lambda E^c \wedge E^b \wedge E^a
- \frac{1}{2} e^{\frac{1}{12} \phi}
\bar{E} \wedge \gamma_{ab} \gamma_{11} \Lambda E^b \wedge E^a
\wedge E^{11}$.
Provided that we neglect the last term $\int_{M_3} \Omega'$ in the
Wess-Zumino action for the time being, this dual action is nothing
but the standard form of the M2-brane action \cite{Sezgin}.
Thus, we have proved that the super D2-brane action in a type IIA
supergravity background is transformed to the super M2-brane
action with a circular compactified 11th dimension
in eleven dimensional supergravity background through a duality
transformation of the world-volume gauge field as expected from
IIA/M-duality.

In fact, we can show the rescaling (\ref{4.9}) that was required
arises the well-known relation between the dilaton and the radius
of a compactified circle in the 11th direction as follows.
{}From Eq.(\ref{4.5}) and the rescaling, we have a relation
\begin{eqnarray}
G^{11}_{ij} = e^{-\frac{1}{6} \phi} G_{ij} + e^{\frac{4}{3} \phi}
(\partial_i B + C_i) (\partial_j B + C_j).
\label{4.11}
\end{eqnarray}
Next in order to get the desired relation we have to rewrite
the Einstein metric $G_{ij}$ in terms of the string metric
$G^{(s)}_{ij} \equiv e^{\frac{1}{2} \phi} G_{ij}$ whose result
is given by
\begin{eqnarray}
G^{11}_{ij} = e^{-\frac{2}{3} \phi} G^{(s)}_{ij} + e^{\frac{4}{3} \phi}
(\partial_i B + C_i) (\partial_j B + C_j).
\label{4.12}
\end{eqnarray}
This correctly yields the relation between the 11 dimensional
metric and the 10 dimensional string metric \cite{Witten2}.
Particularly, the coefficient in front of $(\partial B)^2$
gives us the relation that
$R_{11} = e^{\frac{2}{3} \phi}$ \cite{Witten2}, where
$R_{11}$ is the radius of a compactified circle in the 11th direction.

{}Finally, let us comment the term $\int_{M_3} \Omega'$ in the
Wess-Zumino action. Since $\Lambda_{\alpha} \equiv \frac{1}{2}
\partial_{\alpha} \phi$ as defined in Sec.2, this term never
decouples from the theory unless the dilaton is a constant.
This fact is physically reasonable from the fact
that $\Lambda_{\alpha}$ is the dilatino which is a superpartner
of the dilaton.
Thus, at first sight, since the dilaton and the dilatino are
contained, the dual action (\ref{4.10}) appears
more general than the standard M2-brane action formulated
in an 11-dimensional supergravity background whose bosonic
fields are the metric and a 3-form gauge potential.
But this is illusory since it has been already shown \cite{Duff}
that by suitable redefinitions of
the superconnection and parts of the supervielbein the terms
involving the dilatino can be set to zero in 11 dimensions.
Accordingly, we have precisely shown that the dual of
the super D2-brane action can be identified as the M2-brane
action with a circular 11th dimension.

\section{The super D3-brane}

In this section let us show the $SL(2, Z)$ self-duality of the
super D3-brane action in a general type IIB supergravity
background with non-constant dilaton and axion in both
classically and quantum mechanically exact manner.
In the case of non-constant dilaton and axion there appear
several new features as we will discuss in this section.

{}From Eqs.(1), (2) and (3), the super D3-brane action
is given by

\begin{eqnarray}
S &=& S_{DBI} + S_{WZ}, \nonumber \\
S_{DBI} &=& - \int_{M_4}d^4\sigma \sqrt{-\det(G_{ij}+ e^{-\frac{\phi}{2}}
{\cal F}_{ij})}, \nonumber \\
S_{WZ} &=& \int_{M_4}(C_4 + C_2\wedge {\cal F} +
\frac{1}{2}C_0 {\cal F}\wedge {\cal F})
= \int_{M_{4}}\Omega_{4}.
\label{5-1}
\end{eqnarray}
The constraints (12) on the field strengths of NS-NS
and R-R form-potentials are given by

\begin{eqnarray}
H_3 &= & db_2 = ie^{\frac{\phi}{2}}\bar{E}\wedge\hat{E}
\wedge{\cal K}E + \frac{1}{2}e^{\frac{\phi}{2}}
\bar{E} \wedge \gamma_{ab}{\cal K}\Lambda
E^b\wedge E^a,\nonumber \\
R_{(1)} &=& dC_0 = 2e^{-\phi}\bar{E}{\cal E}\Lambda, \nonumber\\
R_{(3)} &=& dC_2 - H_3 C_0
= -ie^{-\frac{\phi}{2}}\bar{E}\wedge\hat{E}\wedge{\cal I}E
+ \frac{1}{2}e^{-\frac{\phi}{2}}\bar{E} \wedge \gamma_{ab}
{\cal I}\Lambda E^b\wedge E^a, \nonumber \\
R_{(5)} &=& dC_4 - H_3 \wedge C_2
= \frac{i}{6}\bar{E}\wedge
\gamma_{abc}{\cal E}E\wedge E^c\wedge E^b\wedge E^a.
\label{5-2}
\end{eqnarray}

Note that in the case of the non-constant dilaton and axion the
Abelian worldvolume gauge field enters into the axion term in
the Wess-Zumino term only through the
gauge invariant and supersymmetric
${\cal F}$. Therefore this term is no longer a topological term and
the symmetry under the constant shift of the axion seems to be
lost. However, as we will see in the next subsection, it is this
form of the Wess-Zumino action that ensures the NS-NS and
the R-R 2-form potentials transform as a doublet under
the $SL(2,R)$ duality transformation.

\subsection{The classical analysis}

In this subsection we show that the super D3-brane action in a
general type IIB supergravity background is classically
self-dual.
For this we first add a Lagrangian multiplier term \cite{Tseytlin}
\begin{eqnarray}
S_{\tilde{H}} = \int_{M_4}d^4\sigma\frac{1}{2}\tilde{H}^{ij}
(F_{ij}-2\partial_{i} A_{j} ),
\label{5-3}
\end{eqnarray}
to the above action (\ref{5-1}) and solve the equation of motion for
$A_i$ by $\tilde{H}^{ij} = \epsilon^{ijkl}\partial
_k B_l$ with a dual vector potential $B_i$. Then substitution of
this solution into the action (\ref{5-1}) leads to an action
 $S'=S_{1}+S_{2D}$, where
\begin{eqnarray}
S_1 &=& \int_{M_4} \left[-d^4\sigma \sqrt{-\det(G_{ij}+ e^{-\frac{\phi}{2}}
{\cal F}_{ij})} + (C_2 + \tilde{F})\wedge {\cal F}
+ \frac{1}{2}C_0 {\cal F}\wedge {\cal F} \right],
\nonumber \\
S_{2D} &=& \int_{M_4}(C_4 + \tilde{F}\wedge b_{2}),
\label{5-4}
\end{eqnarray}
where $\tilde{F}=dB$.

Next we solve the equation of motion for
$F_{ij}$ and substitute its solution into the action $S'$.
Following \cite{Tseytlin}, we perform the calculation in the specific
local Lorentz frame where $G_{ij}=\eta_{ij}=$ diag$(-1,1,1,1)$ and
${\cal F}$ has the following block diagonal form
\begin{equation}
{\cal F}_{ij} =
\left(
\begin{array}{cccc}
0 & {\cal F}_{01} & 0 & 0 \\
-{\cal F}_{01} &0 & 0 & 0 \\
0 & 0 & 0 & {\cal F}_{23} \\
0 & 0 & -{\cal F}_{23} & 0
\end{array}
\right).
\label{5-5}
\end{equation}
Then the action (\ref{5-4}) is written in this local Lorentz
frame as
\begin{eqnarray}
S^{\prime}&=&\int_{M_{4}}d^{4}\sigma [-\sqrt{(1-e^{-\phi}{\cal
F}_{01}^{2}) (1+e^{-\phi}{\cal F}_{23}^{2})} + (C_2 +\tilde{F})_{23}
{\cal F}_{01} + (C_2 +\tilde{F})_{01}{\cal F}_{23}
\nonumber \\
&+& C_{0}{\cal F}_{01}{\cal F}_{23}
+ C_{0123} + \tilde{F}_{01}b_{23} +\tilde{F}_{23}b_{01} ].
\label{5-6}
\end{eqnarray}
Solving the equation of motion for $F_{01}$ and $F_{23}$ yields
\begin{eqnarray}
{\cal F}_{01} = -\frac{1}{A} \left(e^{\phi}C_{0}b + a \sqrt{\frac
{A-b^{2}}{A+a^{2}}} \right), \
{\cal F}_{23} = -\frac{1}{A} \left(e^{\phi}a - b\sqrt{\frac{A+a^{2}}
{A-b^{2}}} \right),
\label{5-7}
\end{eqnarray}
where
\begin{eqnarray}
A\equiv e^{-\phi}+e^{\phi}C_{0}^{2}, \ a\equiv (C_2 +\tilde{F})_{23}, \
b\equiv (C_2 +\tilde{F})_{01}.
\label{5-8}
\end{eqnarray}
Inserting these solutions into the action (\ref{5-6}) and
returning to the general coordinate reference frame, we arrive
at the dual action $S_{D}$ in a simple form
\begin{eqnarray}
S_D &=& -\int_{M_4}\sqrt{-\det \left[G_{ij} +
\frac{1}{\sqrt{e^{-\phi}+e^{\phi}C_{0}^{2}}}(\tilde{F}_{ij}+C_{2ij})\right]}
+ \int_{M_4}\Omega_D, \nonumber \\
\Omega_D &=& C_4 + b_2\wedge \tilde{F}
 -\frac{1}{2}\frac{e^{\phi}C_{0}}{e^{-\phi}+e^{\phi}C_{0}^{2}}
 (\tilde{F}+C_{2})\wedge (\tilde{F}+C_{2}).
\label{5-9}
\end{eqnarray}

Now if we define a new primed dilaton, axion and form-potentials
which are of the form,

\begin{eqnarray}
e^{-\phi^{\prime}}=\frac{1}{e^{-\phi}+e^{\phi}C_{0}^{2}},
\nonumber \\
C_{0}^{\prime}=\frac{e^{\phi}C_0}{e^{-\phi}+e^{\phi}C_{0}^{2}},
\label{5-10}
\end{eqnarray}
and
\begin{eqnarray}
b_2^{\prime} = -C_2, \ C_2^{\prime} = b_2, \ C_4^{\prime} = C_4 - b_2
\wedge C_2,
\label{5-11}
\end{eqnarray}
then the dual action can be written as
\begin{eqnarray}
S_D &=& -\int_{M_4}d^{4}\sigma\sqrt{-\det(G_{ij}
+ e^{-\frac{\phi^{\prime}}{2}}\tilde{{\cal F}}_{ij}^{\prime})}
\nonumber \\
&{}& + \int_{M_4}(C_4^{\prime} + C_2^{\prime} \wedge
\tilde{{\cal F}}^{\prime} +\frac{1}{2}C_{0}^{\prime}
\tilde{{\cal F}}^{\prime}\wedge\tilde{{\cal F}}^{\prime}),
\label{5-12}
\end{eqnarray}
where $\tilde{{\cal F}}^{\prime}=\tilde{F}-b_{2}^{\prime}$.
The resulting action is completely the same form as the original
action (\ref{5-1}) with which we have started. This means that under
the transformations (\ref{5-10}), (\ref{5-11}) and
\begin{eqnarray}
\tilde{F}\rightarrow F, \
F\rightarrow \tilde{F},
\label{5-13}
\end{eqnarray}
the action (\ref{5-1}) and (\ref{5-12}) are classically equivalent.

In order to establish the $SL(2,R)$ self-duality of the super
D3-brane action in a general type IIB supergravity background,
we have to answer the following two problems.
The first problem is to clarify the transformation
properties of form-potentials  under the
$SL(2,R)$ duality transformation.
The second one is whether the type IIB constraint equations
(\ref{5-2}) are preserved under the $SL(2,R)$ duality
transformation.

First of all, let us introduce complex  variable $\tau$ as in Sec.3,
\begin{eqnarray}
\tau = C_{0} + ie^{-\phi},
\label{5-14}
\end{eqnarray}
then the transformation (\ref{5-10}) is written as $\tau
\rightarrow \tau^{\prime}=-\frac{1}{\tau}$, which is an element
of the $SL(2,R)$ with the transformation matrix
$S=
\left(
\begin{array}{cc}
0 & 1 \\
-1 & 0
\end{array}
\right)$.
According to the transformation rules (\ref{A-10}) and
(\ref{A-11}) the transformations (\ref{5-11}) and
(\ref{5-13}) imply that
$(b_{2}, -C_{2})$ and $(\tilde{F}, -F)$ are expected to form
$SL(2,R)$ doublets which transform like B in (\ref{A-11}) and
A in  (\ref{A-10}), respectively.

Next we must find the $SL(2,R)$ element $\tau \rightarrow
\tau^{\prime}=\tau +1$ which corresponds to the shift of the
axion $C_{0}\rightarrow C_{0}+1$. It is easily shown that
as in the super D-string action  the super D3-brane action
(\ref{5-1}) is invariant under the following two transformations
\begin{eqnarray}
C_{0} &\rightarrow& C_{0}+1, \nonumber \\
b_{2} &\rightarrow& b_{2} , \ C_{2}\rightarrow C_{2} -{\cal F}
=C_{2}+b_{2}-F,
 \nonumber \\
C_{4} &\rightarrow& C_{4} + \frac{1}{2}{\cal F}\wedge {\cal F},
\nonumber \\
F &\rightarrow& F,
\label{5-15}
\end{eqnarray}
and
\begin{eqnarray}
C_{0} &\rightarrow& C_{0}+1, \nonumber \\
b_{2} &\rightarrow& b_{2} , \ C_{2}\rightarrow C_{2}+b_{2},
 \nonumber \\
C_{4} &\rightarrow& C_{4} + \frac{1}{2}b_{2}\wedge b_{2},
\nonumber \\
F &\rightarrow& F.
\label{5-16}
\end{eqnarray}
The super D3-brane action (\ref{5-1}) is exactly invariant
under (\ref{5-15}) and invariant only up to the
topological term $\frac{1}{2}F\wedge F$ under (\ref{5-16}).
It is also easily shown that the type IIB constraint equations
(\ref{5-2}) are invariant under these transformation.
Thus the transformations (\ref{5-15}) and (\ref{5-16}) are
real symmetries of the theory. These transformations are
also consistent with the postulate that $(b_{2}, -C_{2})$
and $(\tilde{F}, -F)$ form $SL(2,R)$ doublets which transform
like B in (\ref{A-11}) and A in  (\ref{A-10}), respectively.

The transformation (\ref{5-15}) is an exact symmetry of the
theory and seems to be more attractive than another one
(\ref{5-16}). It may be interpreted in such a way that
there may exist a manifestly self-dual formulation of the super
D3-brane which are broken by some gauge fixing to the present
formulation of super D3-brane action.
Unfortunately, however, as discussed in the next subsection,
the transformation (\ref{5-15}) does not satisfy the
consistency condition of the constructive relation (\ref{5-25})
in the next subsection.
Therefore we will not consider this transformation any more
in this paper.

At this point we should stress the following fact. In the case of
constant or vanishing dilaton and axion the following
form is sometimes taken as the Wess-Zumino action;
\begin{eqnarray}
S_{WZ}=\int(C_{4} + C_{2}\wedge{\cal F} +\frac{1}{2} C_0
F\wedge F),
\nonumber
\end{eqnarray}
Although this action is trivially invariant up to the topological
term $\frac{1}{2}F\wedge F$ under the shift of the axion
$C_{0}$, the R-R 2-form potential $C_{2}$ can not transform in
such a way to form $SL(2,R)$ doublet with the NS-NS 2-form
potential $b_{2}$ and preserve the constraint equation on
the R-R 5-form strength $R_{(5)}$.
Therefore with this Wess-Zumino action the $SL(2,R)$
duality symmetry would be lost even in the case of constant or
zero dilaton and axion.

The general $SL(2,R)$ transformations are generated
by $\tau\rightarrow -\frac{1}{\tau}$ and $\tau\rightarrow \tau +1$.
Here we recaptulate the $SL(2,R)$ transformation rule for
various fields in the theory for later use in the next subsection.
Corresponding to the $SL(2,R)$ transformation of the complex
variable $\tau$
\begin{eqnarray}
\tau \rightarrow \tau^{\prime}=\frac{a\tau + b}{c\tau + d}, \ ad-bc=1,
\label{5-17}
\end{eqnarray}
which in terms of $C_{0}$ and $\phi$ is equivalent to
\begin{eqnarray}
C_{0}^{\prime}&=&\frac{(aC_{0}+b)(cC_{0}+d)+ace^{-2\phi}}
{(cC_{0}+d)^2 +c^2 e^{-2\phi}}, \nonumber \\
e^{-\phi^{\prime}}&=&\frac{e^{-\phi}}
{(cC_{0}+d)^2 +c^2 e^{-2\phi}},
\label{5-18}
\end{eqnarray}
we define the matrix $S$ as
\begin{equation}
S =
\left(
\begin{array}{cc}
 a & b \\
 c & d
\end{array}
\right), \
(S^{T})^{-1} =
\left(
\begin{array}{cc}
 d & -c \\
 -b & a
\end{array}
\right).
\label{5-19}
\end{equation}
Then it is expected that $\tilde{F},
F, b_{2}, C_{2}$ and $C_{4}$ transform in the following manner;
\begin{eqnarray}
\tilde{F}\rightarrow \tilde{F}^{\prime}=a\tilde{F}-bF, \
F\rightarrow -c\tilde{F} + dF,
\label{5-20}
\end{eqnarray}
\begin{eqnarray}
b_{2}\rightarrow b_{2}^{\prime}=db_{2}+cC_{2}, \
C_{2}\rightarrow C_{2}^{\prime}=bb_{2}+aC_{2},
\label{5-21}
\end{eqnarray}
and
\begin{eqnarray}
C_{4}\rightarrow C_{4}^{\prime}=C_{4}+\frac{bd}{2}
b_{2}\wedge b_{2} + bcb_{2}\wedge C_{2}
+\frac{ac}{2}C_{2}\wedge C_{2}.
\label{5-22}
\end{eqnarray}
The transformation of $C_{4}$ is determined by the requirement
that $R_{(5)}$ is invariant under the $SL(2,R)$
transformations (\ref{5-16}) and (\ref{5-20}). The $SL(2,R)$
transformations (\ref{5-17})-(\ref{5-22}) are just the same
as those in \cite{GG}, but the authors of \cite{GG} gave neither
the transformation of $C_{4}$ nor $SO(2)$ spinor-rotation
(\ref{A-16}) since they considered only the bosonic D3-brane action.
The $SL(2,R)$
transformation (\ref{5-22}) of $C_{4}$ and the $SO(2)$
spinor-rotaition (\ref{A-16}) are indispensable ingredients
for
the type IIB supergravity constraints to be invariant under
the $SL(2,R)$ duality transformation.

 Next we turn to the problem of the invariance (or covariance)
of the type IIB supergravity constraints under the $SL(2,R)$
duality transformation. In Appendix it is shown that
all the type IIB supergravity constraint equations
(of course including the torsion constraints) are
invariant under general $SL(2,R)$ duality transformations
(\ref{5-17}), (\ref{5-21}), (\ref{5-22}) and the
$SO(2)$ spinor-rotation (\ref{A-16}).

Finally we note that, since the invariance under $\tau\rightarrow
\tau + 1$ holds only up to the topological term
$\frac{1}{2}F\wedge F$, the $SL(2,R)$ invariance of the theory
would be broken to the $SL(2,Z)$ invariance in a quantum
theory.

\subsection{The exact analysis}
Let us turn to the other analysis of duality condition which
was initiated by Gaillard and Zumino \cite{GZ1} and extended
by several authors \cite{GZ2,GR,GG,IIK1}.

 Let us define the dual field strength $K_{ij}$ by the
following constructive relation;
\begin{eqnarray}
\ast K^{ij} = \frac{\partial L}{\partial F_{ij}}, \
\frac{\partial F_{kl}}{\partial F_{ij}}= \delta_{k}^{i}
\delta_{l}^{j}-\delta_{l}^{i}
\delta_{k}^{j},
\label{5-25}
\end{eqnarray}
where the Hodge dual components $\ast K_{ij}$ for the antisymmetric
tensor $K_{ij}$ are defined by
\begin{eqnarray}
\ast K_{ij} \equiv
\frac{1}{2}\epsilon_{ij} ^{kl} K_{kl}, \
\ast\ast K_{ij}=-K_{ij},
\label{5-26}
\end{eqnarray}
where $\epsilon_{ijkl}$ is the Levi-Civita symbol in 4 dimensions,
$\epsilon^{0123}=1$ and the signature of $G_{ij}$ is $(-,+,+,+)$.
At the classical level $K_{ij}$ introduced here is equivalent
to the dual variable $-\tilde{F}_{ij}$ in the last subsection.
Therefore it is natural to assume that $(K_{ij}, F_{ij})$
transforms as Eq.(\ref{5-20}); for the infinitesimal
transformation with
$S=
\left(
\begin{array}{cc}
1+\alpha & \beta \\
\gamma & 1-\alpha
\end{array}
\right)$
it is given by
\begin{eqnarray}
\delta K_{ij}= +\alpha K_{ij} + \beta F_{ij}, \
\delta F_{ij}= -\alpha F_{ij} + \gamma K_{ij},
\label{5-27}
\end{eqnarray}

The sufficient condition for the consistency of the constructive
relation  (\ref{5-25}), the invariance of the field equation
and the invariance of the (world-volume) energy-momentum
tensor under the $SL(2,R)$ duality transformation is that
the following equation is satisfied for arbitrary $SL(2,R)$
parameters $\alpha, \beta$ and $\gamma$;
\begin{eqnarray}
\frac{\gamma}{4}*K^{ij}K_{ij}
-\frac{\beta}{4}*F^{ij}F_{ij}
-\frac{\alpha}{2}*K^{ij}F_{ij} + \delta_{\Phi} L =0,
\label{5-28}
\end{eqnarray}
where $\delta_{\Phi}$ means to take the $SL(2,R)$ variation for
all fields except the world volume Abelian gauge field.
This equation is derived by the similar argument in \cite{IIK1}
where a similar equation was derived for the special case
of $\alpha=0$ and $\beta=-\gamma$ ($SO(2)$ duality
transformation). We call Eq.(\ref{5-28}) the Gaillard-Zumino (G-Z)
duality condition.

In \cite{IIK1} it has been shown that the G-Z condition
is actually the necessary and sufficient condition in order
that one can define off-shell (non-local) duality
transformation for the $U(1)$ gauge potential itself under which
the action is invariant up to some surface terms. Therefore
if one can show for an action to satisfiy the G-Z condition,
then one establishes the exact self-duality of the theory
described by this action without resort to any classical
approximation. It has been shown that the super
D3-brane action on a flat \cite{IIK2}, an $AdS_5 \times S^5$
\cite{Kimura} and a general type IIB supergravity backgrounds
with constant dilaton and axion \cite{KO} indeed satisfies the G-Z
condition.

In this subsection we show that the D3-brane action on the most
general type IIB supergravity background described by the action
(\ref{5-1}) and constraints (\ref{5-2}) satisfies the
G-Z duality condition under the $SL(2,R)$ duality transformations
(\ref{5-17})-(\ref{5-22}) and the $SO(2)$ spinor-rotation
(\ref{A-16}).

Before we enter the detailed discussions we should note that
the transformation (\ref{5-15}) does not satisfiy the
consistency condition for the constructive relation (\ref{5-25}).
The consistency condition of the constructive relation demands
the equation
\begin{eqnarray}
\delta *K^{ij}=
\frac{1}{2}\frac{\partial^{2}L}
{\partial F_{kl}\partial F_{ij}}\delta F_{kl}
+ \frac{\partial^{2}L}{\partial \Phi \partial F_{ij}}\delta
\Phi,
\nonumber
\end{eqnarray}
to be satisfied. A simple calculation leads us to a contradictory
result; $*F^{ij}=0$.
It seems to be curious that the exact symmetry of Lagrangian does not
satisfy the consistency condition.

The super D3-brane Lagrangian (\ref{5-1}) in terms of
component fields is written as
\begin{eqnarray}
L = L_{DBI} + \epsilon^{ijkl}(\frac{1}{24}C_{ijkl} +
\frac{1}{4}C_{ij}{\cal F}_{kl} + \frac{1}{8}C_{0}
{\cal F}_{ij}{\cal F}_{kl}),
\label{5-29}
\end{eqnarray}
\begin{eqnarray}
L_{DBI}&=& -\sqrt{-\det(G_{ij} + e^{-\frac{\phi}{2}}{\cal F}_{ij})}
\nonumber \\
&=& -\sqrt{-G}\sqrt{1 +
\frac{e^{-\phi}}{2}{\cal F}_{ij}{\cal F}^{ij}
- \frac{e^{-2\phi}}{16}({\cal F}_{ij}\ast {\cal F}^{ij})^{2}},
\label{5-30}
\end{eqnarray}
where $G = \det G_{ij}$.
Then the dual field strength defined by (\ref{5-25}) is given by
\begin{eqnarray}
*K^{ij}&=& \frac{\partial L_{DBI}}{\partial F_{ij}}
+ *C^{ij} + C_{0}*{\cal F}^{ij},
\nonumber \\
K_{ij}&=&-(*\frac{\partial L_{DBI}}{\partial F})_{ij} + C_{ij}
+ C_{0}{\cal F}_{ij}.
\label{5-31}
\end{eqnarray}

 The infinitesimal $SL(2,R)$ transformations with
$S=
\left(
\begin{array}{cc}
1+\alpha & \beta \\
\gamma & 1-\alpha
\end{array}
\right)$
of various fields in our theory are given by
\begin{eqnarray}
\delta C_{0}&=&2\alpha C_{0} + \beta -
\gamma(C_{0}^2 - e^{-2\phi}), \nonumber \\
\delta \phi &=& 2\gamma C_{0} -2\alpha,
\label{5-32}
\end{eqnarray}
\begin{eqnarray}
\delta K_{ij}= +\alpha K_{ij} + \beta F_{ij}, \
\delta F_{ij}= -\alpha F_{ij} + \gamma K_{ij},
\label{5-33}
\end{eqnarray}
\begin{eqnarray}
\delta b_{ij}= -\alpha b_{ij} + \gamma C_{ij}, \
\delta C_{ij}= +\alpha C_{ij} + \beta b_{ij},
\label{5-34}
\end{eqnarray}
and
\begin{eqnarray}
\delta{C_{4}}=\frac{\beta}{2}b_{2}\wedge b_{2} +
\frac{\gamma}{2} C_{2}\wedge C_{2}.
\label{5-35}
\end{eqnarray}
The infinitesimal $SO(2)$ spinor-rotation is given by
\begin{eqnarray}
\delta \theta&=&-\frac{\gamma e^{-\phi}}{2}{\cal E}\theta, \
\delta \partial_{\alpha}=\frac{\gamma e^{-\phi}}{2}
({\cal E}\partial)_{\alpha}, \nonumber \\
\delta E &=&\frac{\gamma e^{-\phi}}{2}{\cal E}E, \
\delta \bar{E} =-\frac{\gamma e^{-\phi}}{2}\bar{E}{\cal E}.
\label{5-36}
\end{eqnarray}
The infinitesimal $SO(2)$ spinor-rotation (\ref{5-36}) ensures
the invariance of the type IIB supergravity constraints.

Under these infinitesimal $SL(2,R)$ transformations
$\delta_{\Phi}L$ is given by
\begin{eqnarray}
\delta_{\Phi}L
&=& \frac{1}{2}\frac{\partial L}{\partial b_{ij}}\delta b_{ij}
+ \frac{1}{2}\frac{\partial L}{\partial C_{ij}}\delta C_{ij}
+ \frac{\partial L}{\partial\phi}\delta \phi
+ \frac{\partial L}{\partial C_{0}}\delta C_{0}
+ \frac{1}{24}\frac{\partial L}{\partial C_{ijkl}}\delta C_{ijkl}.
\label{5-37}
\end{eqnarray}

The G-Z duality condition demands that
Eq.(\ref{5-28}) must hold for arbitrary variations $\alpha,
\beta$ and $\gamma$. Therefore  coefficients of $\alpha, \beta$
and $\gamma$ should identically vanish, respectively.
Substituting (\ref{5-37}) with (\ref{5-32})-(\ref{5-35}) into
(\ref{5-27}) we obtain following three equatios for $\alpha, \beta,
\gamma$ coefficients;
\begin{eqnarray}
-\frac{1}{4}*K^{ij}K_{ij}-\frac{1}{2}\frac{\partial L}
{\partial b_{ij}}b_{ij}+\frac{1}{2}\frac{\partial L}
{\partial C_{ij}}C_{ij}
-2\frac{\partial L}{\partial \phi}
+2C_{0}\frac{\partial L}{\partial C_{0}}=0,
\label{5-38}
\end{eqnarray}
\begin{eqnarray}
-\frac{1}{4}*F^{ij}F_{ij}
+\frac{1}{2}\frac{\partial L}
{\partial C_{ij}}b_{ij}+\frac{\partial L}
{\partial C_{0}}
+\frac{1}{8}\epsilon^{ijkl}b_{ij}b_{kl}=0,
\label{5-39}
\end{eqnarray}
and
\begin{eqnarray}
\frac{1}{4}*K^{ij}K_{ij}
+\frac{1}{2}\frac{\partial L}
{\partial b_{ij}}C_{ij}
+2C_{0}\frac{\partial L}
{\partial \phi}
-(C_{0}^{2}-e^{-2\phi})\frac{\partial L}
{\partial C_{0}}
+\frac{1}{8}\epsilon^{ijkl}C_{ij}C_{kl}=0,
\label{5-40}
\end{eqnarray}
respectively.

Eq.(\ref{5-39}) ($\beta$ coefficient) is almost
trivially satisfied.
Eq.(\ref{5-38}) ($\alpha$ coefficient) is also easily shown
to  be satisfied, using the following identity
\begin{eqnarray}
\frac{\partial L_{DBI}}{\partial \phi}=-\frac{1}{4}
\frac{\partial L_{DBI}}{\partial F_{ij}}{\cal F}_{ij},
\label{5-41}
\end{eqnarray}
which is derived by the fact that the dilaton $\phi$ is contained
in $L_{DBI}$ in the form of $e^{-\frac{\phi}{2}}{\cal F}$.

The $\gamma$ coefficient (\ref{5-40}) is reduced after
straightforward but somewhat lengthy calculations to the following
equation
\begin{eqnarray}
\frac{1}{2}\epsilon^{ijkl}(\frac{\partial L_{DBI}}{\partial F_{ij}}
\frac{\partial L_{DBI}}{\partial F_{kl}}
+ e^{-2\phi}{\cal F}_{ij}{\cal F}_{kl}) = 0.
\label{5-42}
\end{eqnarray}
This is the only equation which explicitly depends on the
specific form of $L_{DBI}$.
Using the explicit expression (\ref{5-30}) for $L_{DBI}$, it is
easily shown that the equation (\ref{5-41}) is satisfied.
Therefore we have shown that the super D3-brane action
(\ref{5-1}) in the most general type IIB supergravity
background indeed satisfies the Gaillard-Zumino duality
condition.

In \cite{IIK2} it has been shown that the super D3-brane action
in a flat background with constant dilaton and axion which
satisfies the Gaillard-Zumino  condition is pseudo-invariant
under the $SL(2,R)$ (non-local) duality transformation of
the world-volume Abelian gauge field  in both the Lagrangian
and the Hamiltonian formalism. According to the theorem proved
in \cite{IIK1} our above result strongly suggests that
the super D3-brane action in the most general type IIB
supergravity background is also exactly self-dual without resort
to any semi-classical approximation. It would be an interesting
problem to establish it in both the Lagrangian and the Hamiltonian
formalism.
\section{The super D4-brane}

In this section let us start with the super D4-brane action and
perform a duality transformation of the world-volume gauge field
to reach the action obtained by the double-dimensional reduction
of the super M5-brane \cite{Aganagic3, Mario}. The method we
consider is analogous to the one adopted in Section 4, so we shall
follow a similar path of argument as in the super D2-brane.
Like the super D2-brane, the analysis in this section is
purely $\it{classical}$.

This time, from (\ref{2.1}) and (\ref{2.2}) the super D4-brane action
with a Lagrange multiplier term becomes
\begin{eqnarray}
S &=& S_{DBI} + S_{WZ} + S_{\tilde{H}}, \nn\\
S_{DBI} &=& - \int_{M_5} d^5 \sigma e^{\frac{1}{2} \phi}
\sqrt{- \det ( G_{ij} + e^{-\frac{1}{2} \phi}{\cal F}_{ij} )}, \nn\\
S_{WZ} &=& \int_{M_5 = \partial M_6} ( C_5 +
C_3 \wedge {\cal F} + \frac{1}{2} C_1 \wedge
{\cal F} \wedge {\cal F}) = \int_{M_6} I_6, \nn\\
S_{\tilde{H}} &=& \int_{M_5} d^5 \sigma \frac{1}{2}
\tilde{H}^{ij} ( F_{ij} - 2 \partial_i A_j ).
\label{6.1}
\end{eqnarray}
And the constraints (\ref{2.11}) for type IIA on the field strengths
are given by
\begin{eqnarray}
H_3 &=& db_2 = i e^{\frac{1}{2} \phi} \bar{E} \wedge \gamma_{11}
\hat{E} \wedge E + \frac{1}{2} e^{\frac{1}{2} \phi} \bar{E} \wedge
\gamma_{ab} \gamma_{11} \Lambda E^b \wedge E^a, \nn\\
R_{(2)} &=& dC_1 = i e^{-\frac{3}{4} \phi} \bar{E} \wedge \gamma_{11} E
- \frac{3}{2} e^{-\frac{3}{4} \phi} \bar{E} \wedge \gamma_a \gamma_{11}
\Lambda E^a, \nn\\
R_{(4)} &=& dC_3 + H_3 \wedge C_1 \nn\\
&=& \frac{i}{2} e^{-\frac{1}{4} \phi} \bar{E} \wedge
\gamma_{ab} E \wedge E^b \wedge E^a
+ \frac{1}{12} e^{-\frac{1}{4} \phi} \bar{E} \wedge \gamma_{abc}
\Lambda E^c \wedge E^b \wedge E^a, \nn\\
R_{(6)} &=& dC_5 + H_3 \wedge C_3 \nn\\
&=& \frac{i}{24} e^{\frac{1}{4} \phi} \bar{E} \wedge
\gamma_{abcd} \gamma_{11} E \wedge E^d \wedge E^c \wedge E^b
\wedge E^a \nn\\
&{}& + \frac{1}{240} e^{\frac{1}{4} \phi} \bar{E} \wedge
\gamma_{a_1 \ldots a_5} \gamma_{11} \Lambda E^{a_5}
\wedge \ldots \wedge E^{a_1},
\label{6.2}
\end{eqnarray}
from which $C_5$, $C_3$ and $C_1$ are determined.

As usual, we take the variation with
respect to $A_i$, which gives rise to the solution $\tilde{H}^{ij}
= \frac{1}{6} \epsilon^{ijklm} K_{klm}$ with $K = d B$ with $B$ being
a second rank tensor superfield.
After substituting this solution into the action, we obtain the action
$S' = S_1 + S_{2D}$ where $S_1$ and $S_{2D}$ are defined as
\begin{eqnarray}
S_1 &=& - \int_{M_5} d^5 \sigma e^{\frac{1}{2} \phi}
\sqrt{- \det ( G_{ij} + e^{-\frac{1}{2} \phi} {\cal F}_{ij} )}
+ \int_{M_5} ( {\cal H} \wedge {\cal F} + \frac{1}{2} C_1 \wedge
{\cal F} \wedge {\cal F}), \nn\\
S_{2D} &=& \int_{M_5} ( C_5 + K \wedge b_2 ),
\label{6.3}
\end{eqnarray}
with ${\cal H} = K + C_3$. As before, $S_{2D}$ is unaffected under
a duality transformation. Therefore a duality transformation amounts
to solving the equation of motion for $F_{ij}$ in $S_1$ in order to
rewrite the action in terms of $B$ (or its field strength $K$)
instead of $F_{ij}$.
Following the formula in ref.\cite{Aganagic2}, it is tedious but
straightforward to derive the dual action $S_D = S_{1D} + S_{2D}$
where $S_{1D}$ is given by
\begin{eqnarray}
S_{1D} &=& - \int_{M_5} d^5 \sigma
\left[ e^{\frac{1}{2} \phi} \sqrt{-G} \sqrt{1 + z_1 + \frac{z_1^2}{2}
- z_2} - \frac{e^{\phi}}{8(1 + e^{\phi} C_1^2)} \epsilon_{ijklm}
C^i \tilde{{\cal H}}^{jk} \tilde{{\cal H}}^{lm} \right],
\label{6.4}
\end{eqnarray}
where
\begin{eqnarray}
z_1 &=& \frac{1}{2(-G)(1 + C_1^2)} tr (\tilde{G}\tilde{{\cal H}}
\tilde{G}\tilde{{\cal H}}), \nn\\
z_2 &=& \frac{1}{4(-G)^2 (1 + C_1^2)^2} tr (\tilde{G}\tilde{{\cal H}}
\tilde{G}\tilde{{\cal H}}\tilde{G}\tilde{{\cal H}}
\tilde{G}\tilde{{\cal H}}), \nn\\
G &=& \det{G_{ij}}, \nn\\
\tilde{G}_{ij} &=& G_{ij} +C_i C_j, \nn\\
\tilde{{\cal H}}^{ij} &=& \frac{1}{6} \epsilon^{ijklm} {\cal H}_{klm}.
\label{6.5}
\end{eqnarray}

Now let us turn our attention to $S_{2D}$. The conditions
(\ref{6.2}) yield the equation
\begin{eqnarray}
d\Omega_D &\equiv& d (C_5 + K \wedge b_2) \nn\\
&=& R_{(6)} + (K + C_3) \wedge H_3 \nn\\
&=& \frac{i}{24} e^{\frac{1}{4} \phi} \bar{E}
\wedge \gamma_{abcd} \gamma_{11} E \wedge E^d \wedge E^c \wedge E^b
\wedge E^a - i e^{\frac{1}{2} \phi} \bar{E} \wedge \gamma_a \gamma_{11} E
\wedge E^a \wedge {\cal H}   \nn\\
&{}& + \frac{1}{240} e^{\frac{1}{4} \phi} \bar{E} \wedge \gamma_{a_1
\ldots a_5} \gamma_{11} \Lambda E^{a_5} \wedge \ldots \wedge E^{a_1}
- \frac{1}{2} e^{\frac{1}{2} \phi} \bar{E} \wedge \gamma_{ab}
\gamma_{11} \Lambda E^b \wedge E^a \wedge {\cal H} \nn\\
&\equiv& d \Omega_1 + d \Omega_2,
\label{6.6}
\end{eqnarray}
where $d \Omega_1$ and $d \Omega_2$ are respectively defined as
the first plus second terms independent of the dilatino $\Lambda$
and the third plus forth terms dependent on the dilatino in the
third line of the above equation.

As a result,  we have the dual action of the super D4-brane in
type IIA supergravity background
\begin{eqnarray}
S_D &=& - \int_{M_5} d^5 \sigma
\left[ e^{\frac{1}{2}\phi} \sqrt{-G} \sqrt{1 + z_1 + \frac{z_1^2}{2}
- z_2} - \frac{e^{\phi}}{8(1 + e^{\phi} C_1^2)} \epsilon_{ijklm}
C^i \tilde{{\cal H}}^{jk} \tilde{{\cal H}}^{lm} \right]  \nn\\
&{}& + \int_{M_5} (\Omega_1 + \Omega_2).
\label{6.7}
\end{eqnarray}

As in the case of the super D2-brane in Sec.4, we can also set
$\int_{M_5} \Omega_2$ to be zero in terms of redefinitions of
the superconnection and the supervielbein.
With this situation, if the dilaton is vanishing, this dual action
of (\ref{6.7}) exactly reduces to our previous action \cite{KO},
while if the dilaton is vanishing and the background is in a flat
Minkowski spacetime, it becomes the action considered in
Ref.\cite{Aganagic2}.
Then it is straightforward to show that
each action is identical to its corresponding action which is obtained
by the double-dimensional reduction of the super M5-brane \cite{Aganagic3,
Mario} after rearranging the constant dilaton factor in a suitable way.
Following the similar line of argument, it turns out that
the double-dimensional reduction of the super M5-brane action
coincides with the dual super D4-brane action even in a
general type IIA supergravity background under consideration
as suggested by the duality between M-theory and IIA superstring
theory.

\section{Discussions}

In this paper, we have studied the properties of a duality
transformation of super Dp-brane actions ($p = 1, 2, 3, 4$)
in a general type II on-shell supergravity background, which
have been constructed in Refs.\cite{Cederwall1, Cederwall2,
Bergshoeff}.
They have already been investigated in the case of a flat background
with the zero or constant dilaton and axion \cite{Aganagic2}
and in a type II on-shell supergravity background with the zero
or constant dilaton and axion \cite{KO}.
Our presentation in this paper is most general
compared to the previous approaches so we believe that we have
succeeded in showing that various duality symmetries in the super
D-brane actions are indeed valid in a general type II supergravity
background geometry.

The main motivation of the present paper was to take account of
the non-constant dilaton and axion superfields in carrying out
a duality transformation of the super D-brane actions.
To the best of our knowledge, this is the first attempt and
consequently gives us several new fruitful results and insights
as shown thus far.
In particular, from the viewpoint of the world-volume field theory
we have succeeded in showing that the dilaton and the axion
as well as the two 2-form gauge fields, those are, the NS-NS
2-form and the R-R 2-form,
are doublets of the $SL(2,R)$ Mobius group whereas the graviton
and the 4-form gauge potential are singlets in the Einstein
metric. These facts are of course familiar to us but have been
so far proved only in the target space formulation while
our presentation is purely from the world-volume field theory.
The main new issue in the theory at hand is the existence of
a non-trivial symmetry corresponding to a shift of the modulus
field $\tau$. The discovery of this symmetry enables us to
prove the expected duality relations of the super D-brane actions
under the $SL(2,R)$ transformation, in particular, the
self-duality of the super D3-brane action.

Moreover, it was shown that the dual actions of the super D2-brane
and D4-brane actions on a IIA supergravity background with
the constant dilaton, respectively, have the standard forms which are
expected from the dimensional reduction of the M2-brane and
the M5-brane actions on the 11 dimensional supergravity
background.
Note that the dilaton and the dilatino appear in 10 dimensional spacetime
through the Kaluza-Klein compacification on a circle from the M2- and
M5-brane actions in an 11 dimensional supergravity background.

Of course, in this paper our considerations have been focused on
only the super D-brane actions \cite{Cederwall1, Cederwall2,
Bergshoeff}, thus a detailed comparison between the formulation at
present and the other approaches treating different forms of the
p-brane actions \cite{TBC, CW} certainly merits
further investigation in trying to clarify a number of
non-perturbative aspects of D-brane theory.

In the last section in a paper \cite{Aganagic2}, it is stated
that "...... For the most part, our analysis has been classical
and limited to flat backgrounds. The results should not depend
on these restrictions, however."
In our pevious paper \cite{KO}, we have relaxed these restrictions to
some extent, while in this paper, we have removed such restrictions
completely for the super D1-brane and D3-brane.
On the other hand, for the super D2-brane and D4-brane we have
removed the restriction of 'flat background', but we have
presented only the classical analysis. This restriction should
be also removed in future.
But as emphasized in the previous paper \cite{KO},
it is not always clear for us whether it is necessary to
remove this restriction or not since the Dp-brane actions
with $p > 1$ are in essence unrenormalizable so these actions
might describe the low energy effective theory of some underlying
renormalizable theory. This problem still deserves further
investigation.

\vs 1
\begin{flushleft}
{\bf Acknowledgement}
\end{flushleft}
One of authors (I.O.) is grateful to M. Tonin for valuable
discussions. His work was supported in part by Grant-Aid for
Scientific Research from Ministry of Education, Science and Culture
No.09740212.

\appendix
\section*{Appendix}
\setcounter{equation}{0}
\renewcommand{\theequation}{A.\arabic{equation}}
\subsection*{SL(2,R) coset construction of dilaton and axion}
In type IIB supergravity the dilaton and the axion  describe
the coordinates of $SL(2,R)/SO(2)$ coset manifold. In this
appendix we briefly review the $SL(2,R)/SO(2)$
coset description of the dilaton and the axion in the type IIB
supergravity.

Let us parametrize the $SL(2,R)/SO(2)$ coset space
by
\begin{equation}
V = e^{\frac{\phi}{2}}
\left(
\begin{array}{cc}
e^{-\phi} & C_{0} \\
0 & 1
\end{array}
\right),
\label{A-1}
\end{equation}
where $\phi$ and $C_{0}$ are the dilaton and the axion fields,
respectively.
Then $V$ transforms under the $SL(2,R)$ transformation in the
following way
\begin{eqnarray}
V\rightarrow V^{\prime} &=&
 e^{\frac{\phi^{\prime}}{2}}
\left(
\begin{array}{cc}
e^{-\phi^{\prime}} & C_{0}^{\prime} \\
0 & 1
\end{array}
\right) =SVO(S)^{-1} \nonumber \\
&=& Se^{\frac{\phi}{2}}
\left(
\begin{array}{cc}
e^{-\phi} & C_{0} \\
0 & 1
\end{array}
\right)
O(S)^{-1},
\label{A-2}
\end{eqnarray}
where $S$ is an $SL(2,R)$ matrix given by
\begin{equation}
S=
\left(
\begin{array}{cc}
a & b \\
c & d
\end{array}
\right), \
(S^{T})^{-1}=
\left(
\begin{array}{cc}
d & -c \\
-b & a
\end{array}
\right), \
ad-bc=1,
\label{A-3}
\end{equation}
and $O(S)^{-1}$ is an $SO(2)$ matrix defined as
\begin{equation}
O(S)^{-1}=
\left(
\begin{array}{cc}
\cos\lambda & \sin\lambda \\
-\sin\lambda & \cos\lambda
\end{array}
\right).
\label{A-4}
\end{equation}
Eqs.(\ref{A-2}) and (\ref{A-3}) determine the transformation
rule of $C_{0}^{\prime}$ and $\phi^{\prime}$, and $\lambda$;
\begin{eqnarray}
C_{0}\rightarrow
C_{0}^{\prime}&=&\frac{(aC_{0}+b)(cC_{0}+d)+ace^{-2\phi}}
{(cC_{0}+d)^2 +c^2 e^{-2\phi}}, \nonumber \\
e^{-\phi}\rightarrow e^{-\phi^{\prime}}&=&\frac{e^{-\phi}}
{(cC_{0}+d)^2 +c^2 e^{-2\phi}},
\label{A-5}
\end{eqnarray}
and
\begin{eqnarray}
\cos\lambda = \frac{cC_{0}+d}{\sqrt{(cC_{0}+d)^{2}+c^{2}
e^{-2\phi}}}, \
\sin\lambda = \frac{ce^{-\phi}}{\sqrt{(cC_{0}+d)^{2}+c^{2}
e^{-2\phi}}}.
\label{A-6}
\end{eqnarray}

When we define the complex variable $\tau$ by
\begin{eqnarray}
\tau = C_{0}+ie^{-\phi},
\label{A-7}
\end{eqnarray}
Eq.(\ref{A-5}) gives the transformation rule for $\tau$
under the $SL(2,R)$ transformation;
\begin{eqnarray}
\tau \rightarrow \tau^{\prime}=\frac{a\tau +b}{c\tau +d}.
\label{A-8}
\end{eqnarray}
By construction $V$ in (\ref{A-1}) is the $SL(2,R)$ matrix
which transforms the origin of the coset space
to the point $(\phi, C_{0})$; equivalently $\tau_{0}=i \rightarrow
\tau= C_{0}+ie^{-\phi}$.

Next we introduce a symmetric $SL(2,R)$ matrix
\begin{equation}
M=VV^{T}
=e^{\phi}
\left(
\begin{array}{cc}
|\tau|^{2} & C_{0} \\
C_{0} & 1
\end{array}
\right),
\label{A-9}
\end{equation}
which transforms covariantly under $SL(2,R)$ transformation
$M \rightarrow M^{\prime}=SMS^{T}$.
Then one can define two types of $SL(2,R)$ doublet
which transform linearly under
$SL(2,R)$ transformation. The one is
$A=
\left(
\begin{array}{c}
A_{1} \\
A_{2}
\end{array}
\right)
$
which
transforms like
\begin{eqnarray}
A\rightarrow A^{\prime}=SA,
\label{A-10}
\end{eqnarray}
and another is
$B=
\left(
\begin{array}{c}
B_{1} \\
B_{2}
\end{array}
\right)
$
which transforms like $(MB)\rightarrow (MB)^{\prime}=S(MB)$,
that is,
\begin{eqnarray}
B\rightarrow B^{\prime}=(S^{T})^{-1}B.
\label{A-11}
\end{eqnarray}

\subsection*{Invariance of type IIB supergravity constrants under
SL(2,R)}
As discussed in the text the NS-NS 2-form $b_{2}$ and the R-R
2-form $C_{2}$ form an $SL(2,R)$ doublet of $B$-type in
(\ref{A-11}) ;
\begin{equation}
\left(
\begin{array}{c}
b_{2}^{\prime} \\
-C_{2}^{\prime}
\end{array}
\right)
= (S^{T})^{-1}
\left(
\begin{array}{c}
b_{1} \\
-C_{2}
\end{array}
\right),
\label{A-12}
\end{equation}
and the R-R 4-form $C_{4}$ transforms under the $SL(2,R)$
transformation like
\begin{eqnarray}
C_{4}\rightarrow C_{4}^{\prime}=C_{4}+\frac{bd}{2}
b_{2}\wedge b_{2} + bcb_{2}\wedge C_{2}
+\frac{ac}{2}C_{2}\wedge C_{2},
\label{A-13}
\end{eqnarray}
where the $SL(2,R)$ matrix $(S^{T})^{-1}$ is given by (\ref{A-3}).
 Then it is explicitly shown after straightforward and somewhat
lengthy calculations that the type IIB supergravity constraints
\begin{eqnarray}
H_3 &= & db_2 = ie^{\frac{\phi}{2}}\bar{E}\wedge\hat{E}
\wedge{\cal K}E + \frac{1}{2}e^{\frac{\phi}{2}}
\bar{E} \wedge \gamma_{ab}{\cal K}\Lambda
E^b\wedge E^a,\nonumber \\
R_{(1)} &=& dC_0 = 2e^{-\phi}\bar{E}{\cal E}\Lambda, \nonumber\\
R_{(3)} &=& dC_2 - H_3 C_0
= -ie^{-\frac{\phi}{2}}\bar{E}\wedge\hat{E}\wedge{\cal I}E
+ \frac{1}{2}e^{-\frac{\phi}{2}}\bar{E} \wedge \gamma_{ab}
{\cal I}\Lambda E^b\wedge E^a, \nonumber \\
R_{(5)} &=& dC_4 - H_3 \wedge C_2
= \frac{i}{6}\bar{E}\wedge
\gamma_{abc}{\cal E}E\wedge E^c\wedge E^b\wedge E^a,
\label{A-14}
\end{eqnarray}
are transformed to
\begin{eqnarray}
H_3^{\prime} &= & db_2^{\prime}
= ie^{\frac{\phi^{\prime}}
{2}}\bar{E^{\prime}}\wedge\hat{E}
\wedge{\cal K}E^{\prime}
+ \frac{1}{2}e^{\frac{\phi^{\prime}}{2}}
\bar{E}^{\prime} \wedge \gamma_{ab}{\cal K}
\Lambda^{\prime}E^b\wedge E^a,\nonumber \\
R_{(1)}^{\prime} &=& dC_{0}^{\prime}
= 2e^{-\phi^{\prime}}\bar{E}^{\prime}{\cal E}
\Lambda^{\prime},
\nonumber\\
R_{(3)}^{\prime} &=& dC_{2}^{\prime} - H_{3}^{\prime}
C_{0}^{\prime}
= -ie^{-\frac{\phi^{\prime}}{2}}\bar{E}^{\prime}
\wedge\hat{E}\wedge{\cal I}E^{\prime}
+ \frac{1}{2}e^{-\frac{\phi^{\prime}}{2}}\bar{E}^{\prime}
\wedge \gamma_{ab}{\cal I}\Lambda^{\prime}E^b\wedge E^a,
\nonumber \\
R_{(5)}^{\prime} &=& dC_{4}^{\prime} - H_{3}^{\prime}
\wedge C_{2}^{\prime}
= \frac{i}{6}\bar{E}^{\prime}\wedge
\gamma_{abc}{\cal E}E^{\prime}\wedge E^c\wedge E^b\wedge E^a,
\label{A-15}
\end{eqnarray}
where $\Lambda_{\alpha^{\prime}}^{\prime}=\frac{1}{2}
\partial_{\alpha^{\prime}}\phi^{\prime}$, and $E^{\prime}$
and $\partial_{\alpha^{\prime}}$ are the $SO(2)$
spinor-rotated $N=2$ supervielbein and supercoordinate
derivative defined by
\begin{eqnarray}
\theta^{\prime}&=&\exp(-\frac{\lambda}{2}{\cal E})\theta, \
\partial_{\alpha^{\prime}}=(\exp(\frac{\lambda}{2}{\cal E})
\partial)
_{\alpha} \nonumber \\
E^{\prime}&=&\exp(\frac{\lambda}{2}{\cal E})E, \
\bar{E}^{\prime}=\bar{E}\exp(-\frac{\lambda}{2}{\cal E}),
\label{A-16}
\end{eqnarray}
where
\begin{eqnarray}
\cos\lambda=
\frac{cC_{0}+d}{\sqrt{(cC_{0}+d)^2 + c^2 e^{-2\phi}}}, \
\sin\lambda=
\frac{ce^{-\phi}}{\sqrt{(cC_{0}+d)^2 + c^2 e^{-2\phi}}}.
\label{A-17}
\end{eqnarray}
Therefore all the type IIB supergravity constraint equations
(of course including the torsion constraints) are
invariant under general $SL(2,R)$ duality transformations
(\ref{A-8}), (\ref{A-12}), (\ref{A-13}) and the $SO(2)$
spinor-rotation (\ref{A-16}).
We note that the rotation angle $\lambda$ is just the same
as given in Eq.(\ref{A-6}) as expected.

\vs 1

\end{document}